\documentclass[twocolumn, 
superscriptaddress,
 amsmath,amssymb,
 aps, physrev,
prb,
]{revtex4-2}

\usepackage{graphicx}
\usepackage{dcolumn}
\usepackage{bm}
\usepackage{textcomp}

\usepackage{xcolor}
\usepackage{hyperref}

\begin{document}

\preprint{APS/123-QED}

\title{\textbf{Metastability and high-$T_{\rm c}$ superconductivity in A15-type ternary hydride YSbH$_{6}$ at moderate pressure}}


\author{Maélie Caussé}%
 \affiliation{%
 Department of Materials Science and Metallurgy, University of Cambridge, 27 Charles Babbage Road, Cambridge, CB30FS, United Kingdom
}%
\affiliation{
 Advanced Institute for Materials Research, Tohoku University, Sendai, 980-8577, Japan
}%
 \author{Kieran Bozier}
\affiliation{%
 Department of Materials Science and Metallurgy, University of Cambridge, 27 Charles Babbage Road, Cambridge, CB30FS, United Kingdom
}%
\author{Peter I. C. Cooke}
\affiliation{%
 Department of Materials Science and Metallurgy, University of Cambridge, 27 Charles Babbage Road, Cambridge, CB30FS, United Kingdom
}%
\author{Stefano Racioppi}
\affiliation{%
 Department of Materials Science and Metallurgy, University of Cambridge, 27 Charles Babbage Road, Cambridge, CB30FS, United Kingdom
}

 \author{Chris J. Pickard}
\affiliation{%
 Department of Materials Science and Metallurgy, University of Cambridge, 27 Charles Babbage Road, Cambridge, CB30FS, United Kingdom
}%
\affiliation{
 Advanced Institute for Materials Research, Tohoku University, Sendai, 980-8577, Japan
}%

\date{\today}

\begin{abstract}


The discovery of high-temperature superconductors remains a central challenge in materials science. Hydrogen-rich compounds are among the most promising candidates, as they can exhibit phonon-mediated superconductivity at elevated critical temperatures, though their stabilization typically requires extreme pressures.  
Here, we report the identification of YSbH$_{6}$ as a promising superconductor by a multi-stage high-throughput screening on ternary A15-type hydrides, followed by a high-throughput computational search of the Y--Sb--H system, accelerated by ephemeral data derived potentials.
The cubic $Pm\Bar{3}$ YSbH$_{6}$ phase exhibits a predicted critical temperature of 118\,K at 50\,GPa, among the highest $T_{\rm c}$ reported to date for an A15-hydride at this pressure. Thermodynamic analysis shows that YSbH$_{6}$ lies $\sim$100\,meV/atom above the convex hull at 50\,GPa, but only 26\,meV/atom above the hull at 120\,GPa, suggesting possible metastability and synthesis at similar high pressure conditions. The phase is dynamically stable over a wide pressure range (20--120\,GPa), displays kinetic stability at 50\,GPa and elastic stability at 20 and 50\,GPa, key ingredients for long-lived metastable behaviour at moderate pressures.  
These results highlight YSbH$_{6}$ as a benchmark case illustrating the balance between high-$T_{\rm c}$ performance and limited thermodynamic stability in ternary hydrides, and underscore the importance of combined dynamic, thermodynamic, kinetic and elastic stability analyses for guiding experimental synthesis of metastable superconductors.  

\end{abstract}


\maketitle

\section{Introduction} 

The quest for room-temperature superconductors has driven intense theoretical and experimental efforts, motivated by their transformative potential for energy transmission, magnetic technologies, and quantum devices. Among the most promising candidates are hydrogen-rich materials, which, under high pressure, can exhibit phonon-mediated superconductivity with high critical temperatures ($T_{\rm c}$), even approaching room temperature, such as in LaH$_{10}$ with a $T_{\rm c}$ of 250--260\,K~\cite{Somayazulu2019,Drozdov2019}. First-principles calculations have played a central role in these discoveries, accurately predicting several compounds later confirmed experimentally~\cite{Wang2012,Duan2014,Peng2017,Liu2017,DiCataldo2021,Zhang2022}.

However, their practical use is constrained by the high pressures required to stabilize them, which can often exceed 200\,GPa. This limitation has directed the focus of the search to ternary hydrides with stabilization pressures within the megabar range (below 100\,GPa)~\cite{DiCataldo2021,Dolui2024,Sanna2024}. Indeed, superconducting ternary hydrides that are thermodynamically stable at high pressures could remain metastable at a lower pressure, and could potentially be \textit{quenched} to ambient pressure. Di Cataldo \textit{et al.} identified LaBH$_{8}$ as a high-$T_{\rm c}$ superconductor thermodynamically stable at pressures above one megabar (100\,GPa), and dynamically stable down to 40\,GPa~\cite{DiCataldo2021}. Liang \textit{et al.} likewise predicted LaBH$_{8}$~\cite{Liang2021} to be a promising system. Similarly, Zhang \textit{et al.} predicted the $Fm$--$3m$ phase of LaBeH$_8$, characterized by a fluorite-type H-Be alloy backbone, to be metastable and superconducting with a $T_{\rm c}$ of 191\,K at 50\,GPa. Later work resulted in the successful experimental synthesis of LaBeH$_{8}$ at 110\,GPa, which remained metastable down to 80\,GPa~\cite{Song2023}. Recently, two independent studies predicted Mg$_{2}$IrH$_{6}$ would exhibit a superconducting transition temperature in the 80-160\,K  range~\cite{Dolui2024,Sanna2024}. Dolui \textit{et al.} reported that the compound was metastable at ambient pressure and proposed a synthesis pathway \textit{via} Mg$_{2}$IrH$_{7}$. This latter phase is thermodynamically stable above 15\,GPa, and becomes thermodynamically unfavourable compared to the superconducting Mg$_2$IrH$_6$ phase at ambient pressure. However, attempts to synthesize this structure have so far been unsuccessful, with the disordered Mg$_{2}$IrH$_{5}$ phase being formed instead. Mg$_{2}$IrH$_{5}$ is isostructural with Mg$_{2}$IrH$_{6}$ but has a 5/6 site occupancy for hydrogen \cite{Hansen2024}.

Quenching has been achieved for a ternary hydride Y$_{3}$Fe$_{4}$H$_{20}$, which was synthesized at 80\,GPa from a Laves phase hydride YFe$_{2}$H$_{7}$, and recovered at ambient pressure and in ambient air for a few hours~\cite{Causse2025a,Causse2025}. However, this success did not translate to superconductivity at ambient pressure, and other predicted metastable structures have not been successful \cite{Hansen2024}. It is considerably more challenging to predict materials that both feature a superconducting transition and that are possible to fabricate and \textit{quench} to ambient pressure.

Recent years have seen remarkable progress in the search for hydrogen-rich superconductors, including ternary hydrides that range from perovskite-like to A15-derived frameworks~\cite{Zurek2017,FloresLivas2020,Pickard2020,Boeri2022,Zhao2023}, which are suggested to combine high-$T_{\rm c}$ behaviour with potentially reduced stabilization pressures, in some cases down to ambient pressure for perovskites~\cite{Wei2023,Li2025,Cerqueira2024,Wines2024,He2023,Du2024,Dangic2024}. 
However, although many of the theoretically reported phases are dynamically stable, their thermodynamic and kinetic stability remains to be confirmed~\cite{Ma2017,Gao2021,Wei2023,Shutov2024,Li2025,Wang2025b,Yang2025,Zhao2025,Zhang2025a,Tsuppayakornaek2025}. 
This raises an open question about the likelihood of experimental realization of these predicted superconductors. As highlighted in the workshop “Challenges in designing room-temperature superconductors” (CDRTS 2025)~\cite{cdrts2025}, dynamic stability alone is an insufficient criteria to guide synthesis; a robust assessment of thermodynamic and kinetic stability is equally crucial. 

Thermodynamic stability corresponds to the system having the lowest Gibbs free energy among all the species existing in the same convex hull (products and reactants) at a given pressure and temperature. It can be evaluated via the Maxwell construction, which measures the distance from the convex hull in free energy–composition space~\cite{Pickard2006}. Structures lying on the hull are thermodynamically stable, whereas those slightly above it may be metastable if they also satisfy dynamic and kinetic criteria. Dynamic stability can be inferred from phonon dispersions and requires the absence of imaginary phonon modes, ensuring that the structure is robust against infinitesimal atomic displacements. Kinetic stability, in contrast, reflects resistance to large-scale structural transformations (\textit{e.g.}, lattice distortions or reconstructive phase transitions), typically governed by activation barriers at finite temperature, which can trap metastable states over long timescales. Elastic stability, as determined by the Born stability criteria \cite{Born1940} derived from the elastic constants, is a key yet often overlooked assessment criterion. It reflects a material’s response to infinitesimal strains and constitutes a necessary, though not sufficient, condition for mechanical stability under finite stresses.
Together, these four criteria determine the likelihood that a theoretically predicted compound can be synthesized and persist experimentally.  

Since metastable states vastly outnumber thermodynamically stable ones, a large pool of energetically plausible metastable high-$T_{\rm c}$ hydrides can be expected. However, only a fraction of these may be experimentally accessible. Incorporating synthetic feasibility into predictive frameworks could therefore significantly improve the realization of theoretically promising candidates. Recent advances in \textit{ab initio} structure prediction, accelerated by machine learning, now enable systematic exploration of vast chemical and structural spaces~\cite{Ferreira2023,Dolui2024}. This perspective is especially relevant for A15-structure hydrides, which have been highlighted as promising candidates for high-temperature superconductivity~\cite{Vetrano1967,Abe2013,Wei2023,Kuzovnikov2024,Deng2025,Li2025,Zhang2025,Zhang2025c}. A15 phases are A$_3$B compounds adopting the cubic $Pm$$\bar{3}$$n$ space group, where A is typically a transition metal and B any other element. Previous studies have predicted several A15-hydrides with theoretically favorable superconducting properties, yet none were found thermodynamically stable at moderate pressures around 50\,GPa.  

In this work, we combine \textit{ab initio} calculations with machine-learning–accelerated chemical exploration to search for high-$T_{\rm c}$ hydrides in the A15 framework. We focused our search at a moderate pressure of 50\,GPa, since pressures below one megabar are experimentally more accessible. Compared to experiments at higher pressures, a larger culet size can be used in the diamond anvil cell, producing a larger sample that is easier to characterize \textit{post mortem}. This pressure range also facilitates the use of pure H$_2$ as a hydrogen source, avoiding the impurities introduced by using ammonia borane. Moreover, moderate pressure synthesis can facilitate \textit{quenching} at ambient pressure, as exemplified by Y$_{3}$Fe$_{4}$H$_{20}$ where the lowest stabilization pressure was around 60\,GPa~\cite{Causse2025}. Therefore, the 50\,GPa pressure range is a promising region to search for potential ambient metastable superconducting hydrides.

We identify YSbH$_{6}$, a cubic phase ($Pm\Bar{3}$), as a promising candidate with a predicted $T_{\rm c}$ of 118\,K at 50\,GPa. This phase lies 26\,meV/atom above the convex hull at 120\,GPa, suggesting possible metastability, though it becomes less favorable at lower pressure (108\,meV/atom at 50\,GPa). While not on the convex hull and thus not thermodynamically stable, the compound satisfies the dynamical stability criteria and illustrates how kinetic factors may enable the experimental synthesis of metastable hydrides. Indeed, for this level of theory, this distance from the hull is low, as exemplified by the case of Anatase and Rutile, two polymorphs of TiO$_2$. These two phases are both observed experimentally, despite Rutile being calculated (using the PBE exchange correlation functional) to be around 24\,meV/atom in enthalpy above the Anatase phase~\cite{Racioppi2025}.

Furthermore, we apply an approach for more accurate $T_{\rm c}$ estimations on coarse electronic $\mathbf{k}$ grids, aiming to improve the reliability of superconductivity predictions in high-throughput studies~\cite{Bozier2025, Koretsune2017, Morice2017}. As illustrated by the broad range of reported $T_{\rm c}$ values for Mg$_{2}$IrH$_{6}$ (80–160\,K)~\cite{Dolui2024,Sanna2024}, conventional methods remain highly sensitive to computational parameters. Our dual focus—accurate superconducting property prediction and rigorous stability assessment—addresses one of the key challenges in the discovery by high-throughput calculations of room-temperature superconductors at moderate pressures.  

\begin{figure}[t!]
\centering
\includegraphics[width=0.4\textwidth]{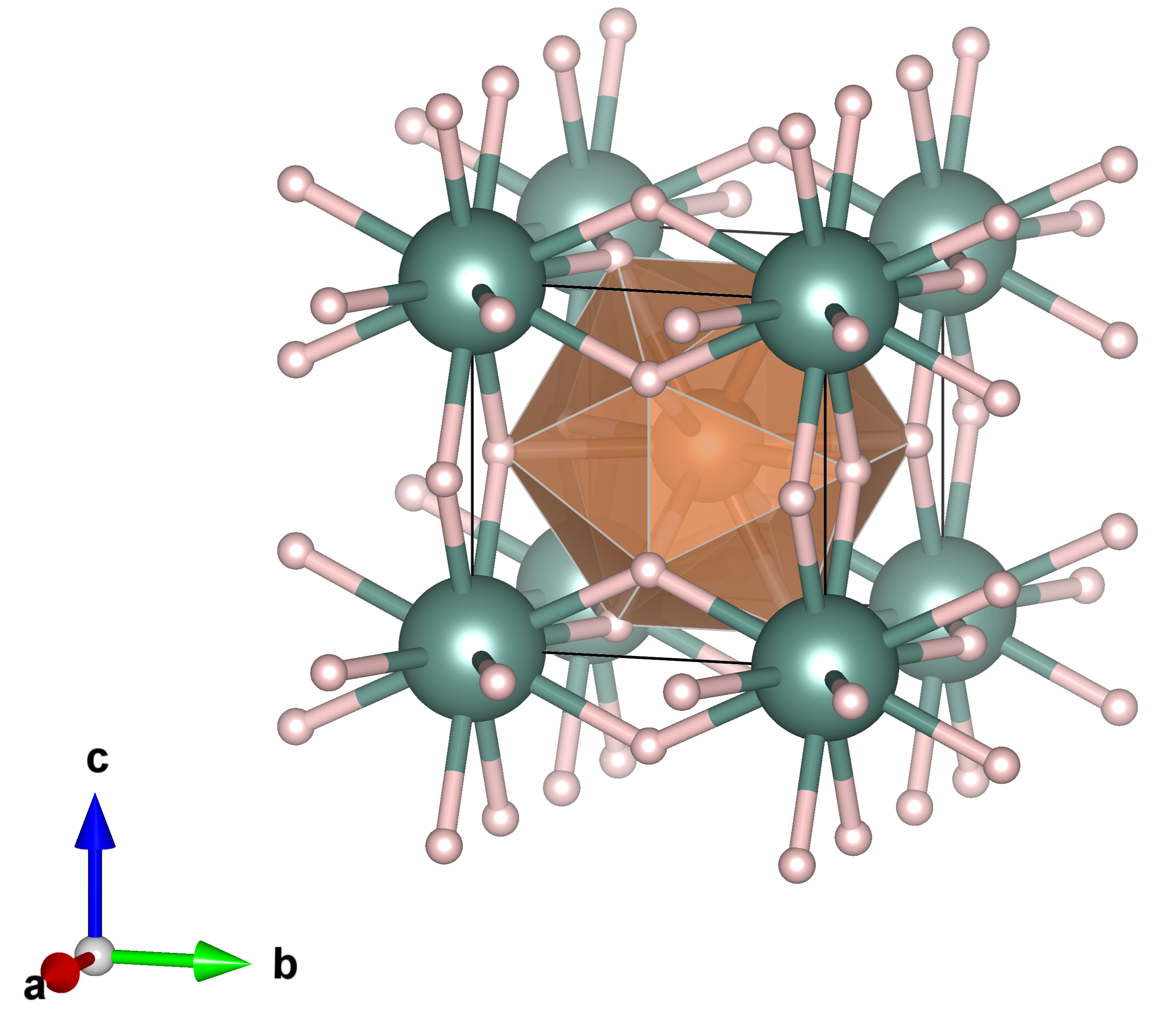}
\caption{The crystal structure of $Pm\Bar{3}$ YSbH$_{6}$. Y, Sb and H atoms are indicated as green, orange and light pink spheres respectively. Y and Sb are surrounded by 12 hydrogen atoms forming a cage. For clarity, only the cage encircling Sb is showcased in orange.}
\label{fig:structure}
\end{figure}

\section{methods} 

We performed an {\it ab initio} prototype search at 50\,GPa on A15-$AB$H$_6$ structures, where $A$ and $B$ were systematically selected from a large subset of the periodic table (see Supplementary Material for details). We combined this constrained structure search with a multi-stage high-throughput screening to find promising high-$T_{\rm c}$ candidates. Using the method introduced in a previous study \cite{Dolui2024}, structures that were non-metallic, displayed magnetic instabilities, or were found to be dynamically unstable (at $\Gamma$) were removed from the search. The superconducting transition temperature of the remaining structures was then calculated and ranked. The highest estimated $T_{\rm c}$ was for $Pm\Bar{3}$-YSbH$_{6}$ ($>$100\,K in the initial search), which motivated the present study and constitutes the main focus of this work.

After confirming dynamic stability at 50\,GPa on $3\times3\times3$ phonon grids in the initial screening, we evaluated the dynamic stability of YSbH$_6$ as a function of applied pressure on finer grids to check for possible charge density wave instabilities. At each pressure, the phonon dispersion and density of states were calculated using \textsc{CASTEP}, with a phonon grid of density $0.06\times2\pi$\,\AA$^{-1}$ and an electronic grid of twice this density. 

To evaluate the superconducting transition temperature $T_{\mathrm{c}}$, we recalculated the phonons on the same $\mathbf{k}$ and $\mathbf{q}$ grids and obtained the electron-phonon matrix elements using the \textsc{Quantum Espresso} code. The matrix elements were evaluated via interpolation onto a fine electronic $\mathbf{k}$ grid with a spacing of $0.015\times2\pi$ \AA$^{-1}$. We used the pseudo-dojo PBEsol \textsc{oncvpsp} potentials \cite{hamann2013, van_setten2018} with a kinetic energy cut-off of 100\,Ry, to ensure that the total energy was converged to within 1 meV/atom. The $T_{\mathrm{c}}$ was then evaluated using both the Allen-Dynes and the isotropic Migdal-Eliashberg equations. A Coulomb pseudopotential of $\mu^{*}_{\mathrm{AD}} = 0.125$ was used in the Allen-Dynes equation. For the Eliashberg equations, the Matsubara cut-off energy was set to 7000\,meV, resulting in a Migdal-Eliashberg Coulomb pseudopotential of $\mu^*_{\mathrm{ME}}=0.227$. The $T_{\mathrm{c}}$ was calculated using both an in-house Eliashberg solver \textsc{a2tc} and the \textsc{IsoME} code~\cite{Kogler2025}. A density of states rescaling factor was applied to the Eliashberg spectral function $\alpha^2F(\omega)$ to accelerate the convergence of $T_{\mathrm{c}}$ \cite{Koretsune2017,Morice2017,Bozier2025}. 

To investigate the overall thermodynamic stability of $Pm\Bar{3}$ YSbH$_{6}$, extensive structure searches on the Y--Sb--H ternary system were performed using Ephemeral Data Derived Potentials (EDDPs)~\cite{Pickard2022, salzbrenner2023}. These machine-learned interatomic potentials were used to accelerate exploration of the energy landscape, and were trained on DFT energies using the same method as in previous studies of the  Mg--Ir--H and Lu--N--H systems~\cite{Dolui2024,Ferreira2023}. We trained several EDDPs tailored for searches at either 50 or 120\,GPa and used these to perform both AIRSS ~\cite{Pickard2006} and hot-AIRSS~\cite{Pickard2025} at these pressures. The anneal was set to last 50\,ps with a temperature range of 1000 to 2000\,K. In total, over 400,000 structures were sampled in our calculations.

To generate the convex-hull of all the structures predicted by EDDPs, we performed DFT geometry optimizations at 0\,K on two collections of around 300 lowest energy structures found at 50 and 120\,GPa respectively. For these well-converged calculations, we used a 900\,eV plane-wave cutoff, a $\mathbf{k}$-point spacing of $2\pi\times 0.02$\,\AA${}^{-1}$, \textsc{CASTEP} C19 pseudopotentials, and the PBEsol exchange--correlation functional. We did not take anharmonicity into account in our calculations.

Further details on the results and elements used in the initial prototype screening, the $T_{\mathrm{c}}$ calculation parameters, the EDDPs training and structure searches, and convex hulls calculated at 50\,GPa are reported in the Supplementary Material.

\section{Results}

We searched for the highest $T_{\rm c}$ hydride within the A15 framework at moderate pressure of 50\,GPa and investigated its synthesizability. In this work, we propose a high-$T_{\rm c}$ ternary hydride, YSbH$_6$, found at 108\,meV/atom above the convex hull at 50 GPa, with an estimated $T_{\rm c} = 118$\,K, as a promising candidate. The structure consists of a cubic $Pm\Bar{3}$ lattice with $a=3.6659$\,\AA~(see Figure~\ref{fig:structure}). The Y and Sb atoms occupy the 1\textit{a} (0, 0, 0) and 1\textit{b} (1/2, 1/2, 1/2) Wyckoff sites respectively, and the hydrogen atoms occupy the 6\textit{g} (0, 0.7425, 0.5) site, forming a polyhedron of 12 hydrogen atoms around Sb and Y (see Fig. \ref{fig:structure}). The metal-hydrogen distances are $d$(Sb--H)$=2.07$\,\AA~and $d$(Y--H)$=2.03$\,\AA~at 50\,GPa. Although the phase remains above the hull at higher pressures, its formation enthalpy decreases to 26\,meV/atom at 120\,GPa, indicating possible metastability. We first combined \textit{ab initio} calculations with exploration of chemical space accelerated by EDDPs to assess thermodynamic stability (section~\ref{thermo}) and to generate molecular dynamic simulations to evaluate the kinetic stability (section~\ref{kinetic}) of YSbH$_{6}$. We then studied the dynamical stability of the structure with first-principles calculations and determined the lowest pressure at which the structure remains dynamically stable (section~\ref{dynamic}). Elastic stability and chemical bonding are discussed in section~\ref{elastic} and \ref{bond}. Finally, two possible synthesis routes are presented in section~\ref{synthesis}.

\subsection{Thermodynamic stability}
\label{thermo}
\begin{figure}[t!]
\centering
\includegraphics[width=0.48\textwidth]{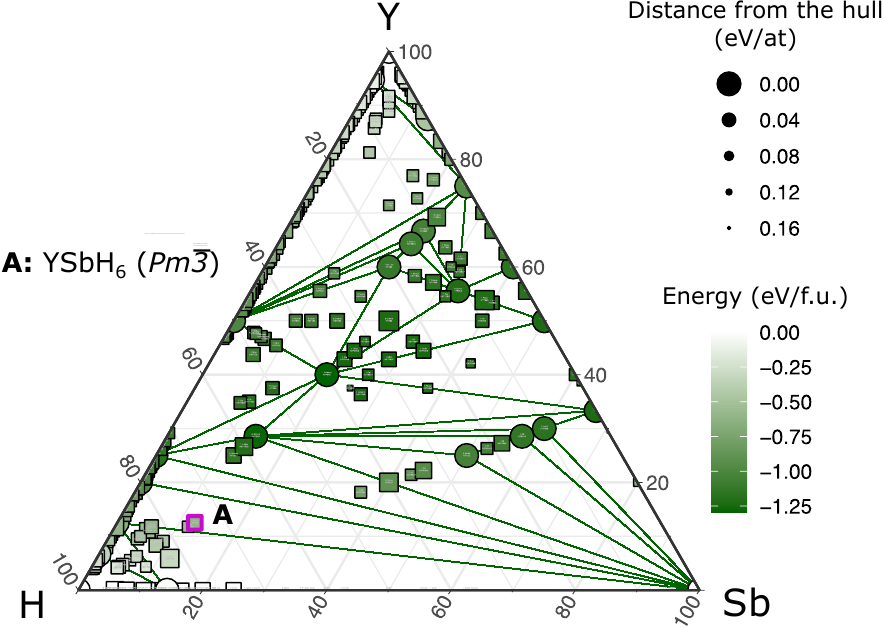}
\caption{Ternary convex hull at 120\,GPa, calculated with the PBEsol functional. Dark green circles indicate the thermodynamically stable phases, solid lines are ridges connecting thermodynamically stable points. Metastable phases are shown as squares, their size is set according the energy distance from the convex hull, the smaller the marker size the larger the distance from the convex hull. The convex hull displays 391 structures collected from different searches, with a total of 311 different compositions. Structures shown on the plot have been relaxed with DFT and showcase an energy distance from the hull E$_h$\,$\leq$\,169\,meV/at. Structures are color-coded according to their formation enthalpy (eV/f.u.) relative to Y, Sb and H elements. At 120\,GPa, YSbH$_{6}$ is 26\,meV/at from the hull and is indicated by a purple square. The list of all stable structures at 120\,GPa is provided in the Supplementary Material (Table ST4) along with the convex hull at 50\,GPa (Fig. S4 and ST5).}
\label{fig:ternary}
\end{figure}

We investigated the synthesizability of YSbH$_6$ by exploring the ternary Y--Sb--H phase diagram at pressures of 50\,GPa and 120\,GPa. We found that although this structure is not on the convex hull at either pressure, its enthalpy distance decreases with applied pressure. Figure~\ref{fig:ternary} shows the resulting convex hull at 120\,GPa, where YSbH$_{6}$ lies 26\,meV/atom above the hull, upon decomposition into Sb, YH$_{4}$ and YH$_{7}$. This small enthalpy difference ($\approx 300$\,K in thermal energy) suggests that the structure may be metastable in the megabar range. The convex hull for 50\,GPa is given in the Supplementary Material, where the structure is found further from the hull ($\approx 100$ meV/atom).

Interestingly, no hydrogen-rich ternary compounds are found on the hull (see the list of structures in Supplementary Material). The highest H/M ratio found in a ternary compound lying on the hull is H/M=4/3 in Y$_{2}$SbH$_{4}$. 
Within our EDDP wide search of the Y--Sb--H system at 120\,GPa, we identified 110 competing phases lying under 26\,meV/atom off the convex hull. Among them, 26 are ternary phases, highlighting the complexity of the Y-Sb-H chemical space. 

Our search at 50 and 120\,GPa are in good agreement with previous theoretical predictions and experimental observations of binary yttrium hydrides \cite{Li2015,Kong2021,Troyan2021,Wang2022}. Cubic YH and YH$_3$ were found on the convex hull at 50\,GPa. Additionally, $I_4/mmm$ YH$_4$ and $C2/c$ YH$_7$ were also found lying on the convex hull. While $Im\Bar{3}m$ YH$_6$ was previously predicted to be on the hull by taking into account zero-point energy (ZPE) and experimentally observed above 100~GPa, the latter was found in our search only 11\,meV/at above the convex hull at 120\,GPa. It highlights the efficiency of using EDDPs to rapidly, thorougly and reliably probe chemical spaces. Built on these results, YSbH$_{6}$ with a distance of 26\,meV above the hull could potentially be synthesised if dynamic, kinetic and elastic stability are confirmed, which we assessed in the following sections.

\subsection{Dynamic stability and $T_{\rm c}$ calculations}
\label{dynamic}

\begin{figure}
    \centering
    \includegraphics[width=0.48\textwidth]{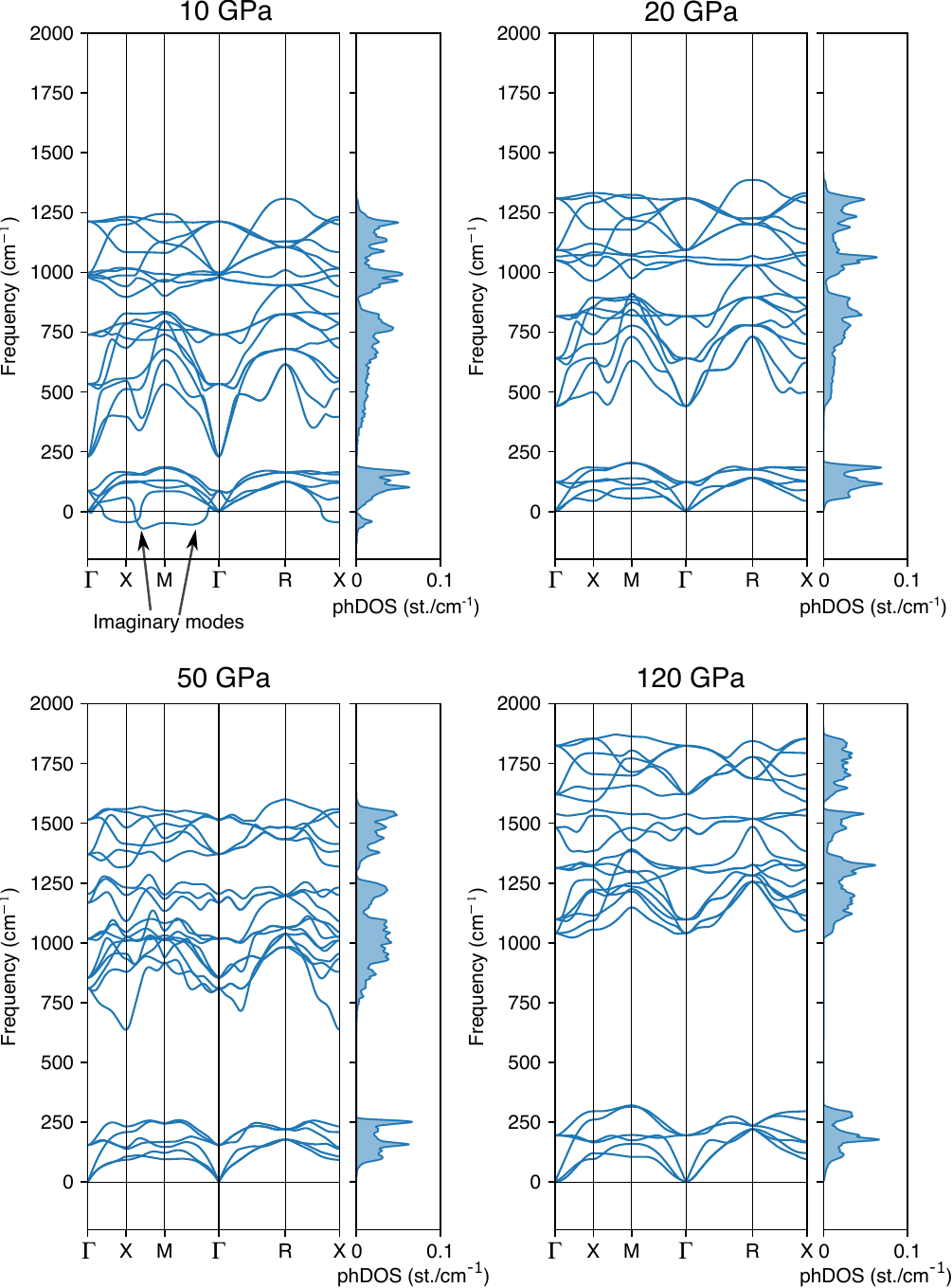}
    \caption{The left panels show the phonon dispersion along a high symmetry path in the Brillouin zone, for different pressures from 10 to 120\,GPa. The $Pm\Bar{3}$ YSbH$_6$ structure is dynamically unstable at 10\,GPa inferred by the presence of imaginary mode, but is stable above 20\,GPa pressure. The right panels show the corresponding phonon density of states (phDOS).}
    \label{fig:castep-phonon-comparison}
\end{figure}

The phonon and electron-phonon properties of $Pm\Bar{3}$ YSbH$_6$ were calculated for pressures from 0 to 120\,GPa. The phonon dispersion relation and density of states are shown in Figure \ref{fig:castep-phonon-comparison}. The phonon frequencies soften as the pressure is lowered, and the structure becomes dynamically unstable below approximately 20\,GPa, as evidenced by the imaginary (negative) frequencies seen in the 10\,GPa results. As this softening can lead to very large electron-phonon coupling strengths ($\lambda$), which can point to a charge density wave instead of superconductivity, we calculated $T_{\mathrm{c}}$ at the lowest dynamically stable pressure (20\,GPa) and further from the structural transition (50\,GPa). Using the PBEsol exchange-correlation functional, the calculated $T_{\mathrm{c}}$ at 50\,GPa ranged from 110--135\,K, depending on the exact method used. A summary of the obtained values is given in Table \ref{tab:Tc_50GPa_pbesol}. A slightly higher $T_{\mathrm{c}}$ was observed for the same pressure if instead the PBE functional was used (135--166\,K). This enhancement is in part due to a slightly higher density of states at the Fermi energy and consequently a higher electron-phonon coupling strength of $\lambda = 2.34$ at 50\,GPa. In this work, we have used the PBEsol exchange-correlation functional unless otherwise stated, as it was specifically revised to be more accurate than PBE for structural properties of bulk solids~\cite{Perdew2008}.

\begin{table}[th]
    \caption{\label{tab:Tc_50GPa_pbesol} Table of the calculated $\lambda$ and $T_{\mathrm{c}}$ using the PBEsol exchange--correlation functional. AD is the Allen-Dynes equation. Both \textsc{a2tc} and \textsc{IsoME} solve the Isotropic Eliashberg equations, in either the constant density of states (cDOS) or variable density of states (vDOS) approximation~\cite{Kogler2025}.}
  \begin{ruledtabular}
     \begin{tabular}{ccccccc}
       Pressure& $\lambda$ & AD & \textsc{a2tc} & \multicolumn{2}{c}{\textsc{IsoME}} \\
       (GPa)  &           &    (K)         &  cDOS (K)       & cDOS (K) & vDOS (K) \\
      \hline
      20 & 3.44 & 86  &  132 & 130 & 114 \\
      30 & 2.21 & 114  & 136  & 134 & 120\\
      40 & 1.95 & 108  & 130 & 130 & 118\\
      45 & 2.07 & 116 &  143 & 143 & 119 \\
      50 & 1.95 & 110 & 135 & 134 & 118 \\
      \end{tabular}
  \end{ruledtabular}
\end{table}

\begin{figure}[h]
\centering
\includegraphics[width=0.48\textwidth]{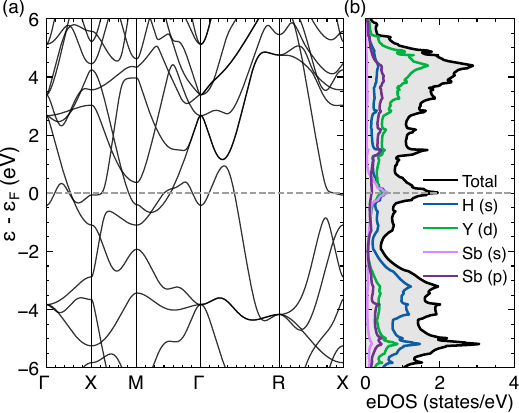}
\caption{(a) Electronic band structure of $Pm\Bar{3}$ YSbH$_6$ at 50\,GPa. (b) The total electron density of states (eDOS) projected onto hydrogen-1\textit{s} (blue), yttrium-4\textit{d} (green), and antimony-5\textit{s} (pink) and antimony-5\textit{p} (purple). H (s), Y (d), Sb (s) and Sb (p) orbitals all make contributions to the peak in the density of states at the Fermi energy. Black dashed line indicates the Fermi level.}
\label{fig:pdos}
\end{figure}

An accurate evaluation of $T_{\mathrm{c}}$ for this structure is complicated by the sharp features in the electronic density of states (DOS). Figure \ref{fig:pdos} shows the electronic band structure together with the total and projected DOS. The flat bands visible along part of the X-M and X-R high-symmetry paths in Figure \ref{fig:pdos} (a) are likely responsible for the van Hove singularity at the Fermi energy seen in Figure \ref{fig:pdos} (b). 
This peak in the density of states is partly responsible for the large electron phonon coupling strength, and consequently the high $T_{\mathrm{c}}$ of this structure. Such a sharply varying DOS near the Fermi energy presents a challenge when using the constant density of states (cDOS) Eliashberg equations. These equations assume that the DOS is constant and fixed to the value at the Fermi level, denoted $N_{\rm{F}}$. Consequently, this method tends to overestimate $T_{\mathrm{c}}$ in systems with a peak at the Fermi level. In comparison, the variable density of states (vDOS) approach is able to account for the lower DOS away from the Fermi energy  \cite{Lucrezi2024, Kogler2025}, resulting in a lower predicted $T_{\mathrm{c}}$ for this system. Therefore, we consider the vDOS values reported in Table \ref{tab:Tc_50GPa_pbesol} to be the most reliable estimates of $T_{\mathrm{c}}$ for this system. The $T_{\rm{c}}$ at 50\,GPa is hence predicted to be 118\,K.

Additionally, since the cDOS $T_{\rm{c}}$ is highly sensitive to the value of $N_F$, small shifts in the Fermi level can lead to large changes in the evaluated transition temperature. These shifts can be achieved by chemical doping or by changes in the applied pressure. Although the Fermi energy at 50\,GPa already lies close to the peak in the DOS, lowering the pressure to 45\,GPa increases the $N_{\rm{F}}$ value to 1.90\,states/eV/cell. As expected, this directly corresponds to a maximum in the cDOS $T_{\rm{c}}$ of 143\,K. However, in comparison the vDOS estimate is significantly less sensitive to the exact peak position.  If the vDOS prediction is more accurate, it suggests $T_{\rm{c}}$ is actually largely insensitive to the pressure tuning, contrary to what might be inferred from the cDOS results alone. A plot of the density of states at different pressures, and the superconducting gap parameter as a function of temperature are shown in Section 2 of the Supplementary Material

Finally, a detailed inspection of the interpolated phonon frequencies around the $\Gamma$ point for the 120\,GPa structure appeared to show a very small imaginary mode at $\mathbf{q} = (0, 0, 0.1)$. We re-evaluated this mode with DFPT using both the standard $10\times10\times10$ and a denser $20\times20\times20$ electronic grid. The coarser grid yielded only positive frequencies, but the finer grid produced a single imaginary mode of frequency -11 cm$^{-1}$. This mode corresponds to a slight distortion involving a shearing motion and a rotation of a hydrogen pair, which breaks the cubic symmetry. The distortion yields a monoclinic structure ($P2/m$ space group) with $\beta = 91.9^{\circ}$ and lattice parameters $a=3.431$\,\AA{}, $b=3.440$\,\AA{}, and $c=3.474$\,\AA{} at 120\,GPa. Enthalpy calculations indicate the distorted structure is $<$1\,meV/atom lower in enthalpy than the cubic structure. ZPE calculations using \textsc{phonopy} on a $3\times3\times3$ phonon grid show the two structures have nearly identical zero-point energies (171 meV/atom  for the cubic and 170 meV/atom for the distorted), again suggesting very little energetic difference between them. As the pressure is lowered, the distorted structure remains slightly energetically favorable, but the magnitude of the distortion decreases and the structure tends towards the cubic phase. The electron-phonon coupling strength in the distorted structure is lower, with $\lambda=1.56$ at 50\,GPa, but the corresponding vDOS critical temperature remains high at around 110\,K (see Section 2 of Supplementary Material for further details).

\subsection{Kinetic stability}
\label{kinetic}

We performed molecular dynamic simulations using EDDPs on a $Pm\Bar{3}$ YSbH$_{6}$ compound at 120\,GPa and 50\,GPa at 300\,K to investigate the  kinetic stability, as was done in Ref.~\cite{Dolui2024}. There is very little diffusion at 300K at both pressures on a long trajectory of 0.5\,ns and the hydrogen cage at both 50 GPa and 120 GPa remained intact up to at least 700 K (see Supplementary Material Section 6). This indicates the cubic YSbH$_{6}$ possesses a high degree of kinetic stability and could thus be metastable and potentially synthesized at high pressure. At 50\,GPa there is very little diffusion, so we can expect the structure to persist after \textit{quenching} from 120 to 50\,GPa.

\subsection{Elastic stability}
\label{elastic}
The elasticity of a lattice is described by its matrix of second-order elastic constants:
\begin{equation}
    C_{ij} = \frac{1}{V_{0}} \frac{\partial^2 E}{\partial \epsilon_i\partial\epsilon_j},
\end{equation}
where $E$ is the energy of the crystal, $V_0$ its equilibrium volume, and $\epsilon_i$ denotes a strain in Voigt notation. $C_{ij}$ is referred to as the stiffness constants, and is a symmetric matrix of dimensions 6$\times$6. 

The ``Born elastic stability criteria"~\cite{Born1940,Mouhat2014} for elastic stability are satisfied if, for a cubic structure,
\begin{equation}
C_{11} - C_{12} > 0,\quad C_{11} + 2C_{12} > 0,\quad\&\quad C_{44} > 0.
\label{eq}
\end{equation}
We evaluated the elastic stability of the $Pm\Bar{3}$ YSbH$_{6}$ structure by calculating the elastic constants $C_{ij}$ at different pressures. 
The cubic phase has a positive $C_{44}$ elastic constant at 20 and 50\,GPa, and fulfils the first two conditions of Eq.~(\ref{eq}), yielding the elastic stability of $Pm\Bar{3}$ YSbH$_{6}$ at these moderate pressures (see Section 2 in the Supplementary Material). However the phase is elastically unstable at high pressure with a negative $C_{44}$ elastic constant at 120\,GPa, which is consistent with the slight phonon instability observed in section~\ref{dynamic}.

\subsection{Chemical bond analysis}
\label{bond}

To gain insight into the chemical bonding of the new A15-type YSbH$_6$, we computed the electron localization functions (ELF) as illustrated in Fig.~S9, and performed the topological analysis of the electron density (Section 7.2 in the Supplementary Material) \cite{Bader1990}. The results show that electrons are primarily localized around the hydrogen atoms, with a clear elongation toward the Sb atoms as seen for AXH$_6$ in ref~\cite{Wei2023}. The charge transfer from Sb and Y to H is confirmed by the Bader integration, which calculate a total charge of -0.35 electrons per hydrogen atom. The ELF values between Sb and H is around 0.60 which emphasizes the presence of weak covalent Sb--H bonds, as also quantified by the topological analysis. As a result, YSbH$_6$ forms SbH$_{12}$ units (see Fig. \ref{fig:structure}). The ELF values near the shortest Y--H separations approach zero (Fig.~S9). Yet, the electron density shows that a Y--H interaction exists, even if weaker than the more covalent Sb--H (Table ST13). Although no significant ELF localization is observed in Fig.~S9, the electron density reveals the presence of H--H interactions; these are, however, weaker than the corresponding Sb--H and Y--H interactions.

\subsection{Possible synthesis routes}
\label{synthesis}

We now discuss the possible synthesis route to achieve $Pm\Bar{3}$ YSbH$_{6}$ under megabar pressure in a diamond anvil cell. Experimentally, YSbH$_6$ could be synthesized via either compression of YSb powder in a hydrogen gas environment, like in Ref.\cite{Causse2025}; or using YH$_3$ powder together with SbH$_3$ gas environment. According to our calculations, both synthesis routes have the potential to access the metastable YSbH$_6$ phase under pressure. Further details on the proposed synthesis pathways are reported in Section 8 in the Supplementary Material. The synthesized phase under pressure could then potentially be \textit{quenched} to lower pressures, given the structure's dynamic, kinetic and elastic stability at 20-50\,GPa.



\section{discussion}

Our results highlight both the promise and the challenges of stabilizing high-$T_{\rm c}$ superconductivity in ternary A15-type hydrides. Although $Pm\Bar{3}$ YSbH$_6$ displays a high predicted $T_{\rm c}$ in the range 110--135\,K at 50\,GPa, its thermodynamic metastability relative to the convex hull (E$_h$=108\,meV/atom at 50\,GPa and 26\,meV/atom at 120\,GPa) suggests that experimental synthesis may only be achievable above the megabar regime. Calculation of formation enthalpy does not take into account zero-point energy, which has been shown to impact the stability of some hydrides. These corrections could stabilize the compound in the same way as LaH$_{10}$ and RbPH$_3$~\cite{Errea2020,Dangic2024}.

The absence of ternary hydrides with high hydrogen content (H/M\,$\geq$\,3) on the convex hull of the Y--Sb--H phase diagram highlights a general trend: hydrogen-rich ternaries are thermodynamically less favorable than their binary counterparts. Compared to the absent ternaries, several binary yttrium hydrides YH, YH$_3$, YH$_4$ and YH$_7$ all lie on the hull, and coincide with theoretical predictions and experimental observations at around 120\,GPa~\cite{Li2015,Kong2021,Troyan2021,Wang2022}. 
Nevertheless, the dynamical stability of YSbH$_6$ down to 20\,GPa, combined with molecular dynamics simulations that indicate kinetic resilience, is encouraging. A previous study explored various A and X elements in the AXH$_{6}$ structures in the pressure range of 50–400 GPa~\cite{Wei2023}. Wei \textit{et al.} kept structures stable with respect to their decomposition into the binary hydrides and the single elements, and seems to have missed YSbH$_6$ as a promising compound.

The harmonic phonon calculations of the $Pm\Bar{3}$ phase at 10\,GPa reveal small phonon instabilities throughout X-M path on the Brillouin zone. Including ionic quantum fluctuations and anharmonicity within the stochastic self-consistent harmonic approximation (SSCHA) as shown for the perovskite RbPH$_3$~\cite{Dangic2024}, the phase might be dynamically stabilized at 10\,GPa or lower pressure.

The calculated superconducting transition temperatures show some variation between the different methods, ranging from 110\,K with the Allen-Dynes up to 135\,K with the constant DOS (cDOS) Eliashberg approach at 50\,GPa. As mentioned earlier, we expect the sharp peak in the DOS at the Fermi energy to cause an overestimation for the cDOS Eliashberg equations, since these instead assume the DOS is constant and fixed to the value at the Fermi level. As such, we suggest that the best estimate is instead given by the variable DOS value of 118\,K, which happens to be very close to the Allen-Dynes prediction. A further source of uncertainty arises from the choice of Gaussian smearing width used in the evaluation of the spectral function $\alpha^2F(\omega)$. Although the rescaling procedure described in Ref.~\cite{Bozier2025} eliminates much of this variation, a spread of about 20\,K remains as the smearing is varied between 0.005\,Ry--0.250\,Ry. We report the $T_{\rm{c}}$ values obtained using a 0.025\,Ry smearing, which provided the most conservative estimates of $T_{\rm{c}}$. A further point of interest was the disparity between PBE and PBEsol functionals in the prediction of the critical temperature value. At 50\,GPa, the superconducting temperature is predicted to be 135--166\,K with PBE, about 25\,K higher than the PBEsol results. We suggest that part of the difference in the predicted $T_{\rm{c}}$ values between the two functionals arises from the higher value of the density of states at the Fermi level ($N_{\rm{F}}$). The value obtained for PBE was $N_{\rm{F}} = 1.97$\,states/eV/cell, while the PBEsol value was $N_{\rm{F}} = 1.76$\,states/eV/cell. This, combined with slightly softer phonons, resulted in a larger electron-phonon coupling of $\lambda = 2.34$ with the PBE functional. Regardless of the exact method used, all predictions of $T_{\rm{c}}$ remained above 100\,K at 50\,GPa, indicating that YSbH$_6$ is a promising candidate for high-temperature superconductivity at moderate pressures.

While EDDP-based molecular dynamics suggest kinetic stability, the predictive accuracy ultimately depends on the quality of the training dataset and would ideally be cross-validated against direct first-principles molecular dynamics. However, thanks to the efficiency of machine-learning potentials, EDDPs enable simulations extending to hundreds of picoseconds, 
a timescale inaccessible to conventional \textit{ab initio} molecular dynamics. Such cross-validation, although desirable, remains prohibitively expensive computationally.

Overall, our findings emphasize the need for combined thermodynamic, dynamical, kinetic, and elastic assessments in the design of new superconductors. YSbH$_6$ stands as an instructive case where high-$T_{\rm c}$ coexists with thermodynamic metastability, underlining that synthesis feasibility rather than $T_{\rm c}$ alone should guide the prioritization of candidate compounds for experimental pursuit.

\vspace{1em}

\section{conclusion}

In summary, we have identified a metastable ternary hydride, YSbH$_{6}$ with A15 symmetry, as a promising high-$T_{\mathrm{c}}$ superconductor at moderate pressures. Using a recently proposed rescaling of the Eliashberg spectral function $\alpha^2F(\omega)$ to accelerate the convergence of the $T_{\rm{c}}$ calculation, we predict a critical temperature of 118\,K. 
A detailed assessment of dynamical, thermodynamic, kinetic and elastic stability, supported by machine-learning potentials, indicates that YSbH$_{6}$ can likely be stabilized at 120\,GPa and may persist as a metastable phase at lower pressures 20 and 50\,GPa. 
These results establish YSbH$_{6}$ as a representative case that illustrates the balance between enhanced superconducting performance and limited thermodynamic stability in ternary hydrides, and they highlight the necessity of combining dynamical, thermodynamic, kinetic, and elastic stability analyses to guide experimental efforts toward metastable superconductors.


\textit{Supplementary Material.}
See the Supplementary Material for additional information.

\textit{Acknowledgements.}
We gratefully acknowledge Lewis Conway, Ryuhei Sato and Francesco Belli for their discussion. We thank Tim Kamsma and William Galloway for their insightful comments on the manuscript. 
K.B. acknowledges funding from an ESPRC DTP studentship in the Department of Materials Science and Metallurgy. This work was supported by the Advanced Institute for Materials Research at Tohoku University (M.C.), the Deep Science Fund at Intellectual Ventures (P.I.C.C.) and Theoretical Condensed Matter Cambridge, Critical Mass Grant - Grant Ref: EP/V062654/1 (S.R.).

\textit{Authors contribution.}

M. C., K. B. and C. J. P. conceptualized the work. M.C. performed the thermodynamic stability calculations, formation enthalpies, ELF and elastic constants calculations. C. J. P. performed the prototype structure search and screening. K.B. performed the dynamic stability, superconductivity and ZPE calculations. S.R. performed the enthalpy calculations for the cubic and distorted phases, and the topological analysis of the electron density. P.I.C.C. performed the molecular dynamics and kinetic stability study. M. C. wrote the original draft, and all authors contributed to the manuscript.

\bibliography{biblio}
\clearpage 

\end{document}


\maketitle
\large

\setcounter{tocdepth}{1}
\tableofcontents
\newpage

\section{Computational Methods}

\subsection{A15--ABH$_6$ high throughput screening}

A high-throughput screening was performed on A15--ABH$_6$ structure (atomic positions described in Table~\ref{tab:abh6cell}, where A and B were randomly chosen from a palette of the periodic table. The palette contained; H, He, Li, Be, B, C, N, O, F, Ne, Na, Mg, Al, Si, P, S, Cl, Ar, K, Ca, Sc, Ti, V, Cr, Mn, Fe, Co, Ni, Cu, Zn, Ga, Ge, As, Se, Br, Kr, Rb, Sr, Y, Zr, Nb, Mo, Tc, Ru, Rh, Pd, Ag, Cd, In, Sn, Sb, Te, I, Xe, Cs, Ba, La, Ce, Yb, Lu, Hf, Ta, W, Re, Os, Ir, Pt, Au, Hg, Tl, Pb, Bi, Po, At.)

First, 5476 structures were generated and successfully relaxed with coarse DFT parameters. Geometry optimizations were performed using \textsc{CASTEP}~\cite{Clark2005} with Perdew-Burke-Ernzerhof exchange--correlation functional~\cite{Perdew1996}, \textsc{CASTEP} QC5 pseudopotentials, a 340\,eV plane-wave cutoff, and a $k$-point spacing of $2\pi\times~0.07$\,\AA${}^{-1}$. We then performed a spin-polarized DFT calculation and removed any structures with non-zero spin, reducing the dataset to 966 structures. For assessing dynamical stability, we performed two iterations of \textsc{CASTEP} phonon calculations, first on a $1\times1\times1$ $q$-grid  (instabilities on the $\Gamma$ point) and then $2\times2\times2$ $q$-grid, removing structures with imaginary phonon modes, reducing the dataset to 823 and then to 375 structures. We calculated the electronic densities of states and eliminated all non-metallic structures reducing the dataset to 131 structures (5 calculations did not converged). Finally, this procedure resulted in a subset of 126 structures for which the $T_{\rm c}$ was estimated by calculating the electron--phonon interactions within density functional perturbation theory (DFPT) using \textsc{Quantum Espresso (QE)}~\cite{Giannozzi2020}, with a set of coarse parameters, and solving the isotropic Migdal-Eliashberg (ME) equations for the Eliashberg functions, $\alpha^2 F(\omega)$. Of the remaining 126 cubic structures, the highest $T_{\rm c}$ calculations are shown in Table~\ref{best-screening}.

\begin{table}[h!]
    \centering
    \caption{ABH$_{6}$ atomic positions}
    \begin{tabular}{llll} \hline \hline
Atom   & x & y& z     \\ \hline
H  & 0.5 &   0 &  0.75\\
H  & 0.25 &   0.5 &  0\\
H  & 0 &   0.75  &  0.5\\
H  & 0.5 &   0  &  0.25\\
H  & 0.75 &   0.5  &  0\\
H  & 0 &   0.25 &  0.5\\
A & 0 &   0  &  0\\
B & 0.5 &   0.5  &  0.5\\ \hline \hline
    \end{tabular}
    \label{tab:abh6cell}
\end{table}

\begin{table}[htbp]
\centering
\caption{High-throughput screening of A15--ABH$_6$ structures with coarse $T_{\rm c}$ calculations. The first ten structures with highest predicted $T_{\rm c}$ are showcased, along with the electron-phonon coupling parameter $\lambda$, phonon logarithmic average frequency $\omega_{log}$ (K) and critical temperature $T_{\rm c}$ from the McMillan equation, the Allen-Dynes modified McMillan equation and calculated via solution of the Eliashberg equations for $Pm\Bar{3}$ AXH$_6$ compounds. $T_{\rm c}$s are given with the electronic Gaussian broadening values 0.05\,Ry.}
\vspace{1em}

\begin{tabular}{lccccccc}
\hline\hline
ahb6-A-B & $\lambda$ & $\omega_{log}$ (K) & $T_{\rm c}^{McM}$(0.05\,Ry) K & $T_{\rm c}^{AD}$(0.05\,Ry) K & $T_{\rm c}^{Eli}$(0.05\,Ry) K  & Formula & spg \\
\hline
abh6-Y-Sb   & 2.202 & 837.555 & 121.668 & 149.554 & 183.097 & YSbH$_6$   & Pm-3 \\
abh6-Sb-Yb  & 8.209 & 190.857 & 43.959  & 124.096 & 151.236 & SbYbH$_6$  & Pm-3 \\
abh6-Bi-Ce  & 2.010 & 699.723 & 95.405  & 115.437 & 147.798 & CeBiH$_6$  & Pm-3 \\
abh6-La-Bi  & 1.782 & 758.938 & 94.152  & 109.711 & 133.532 & LaBiH$_6$  & Pm-3 \\
abh6-Bi-Y   & 1.614 & 844.517 & 95.895  & 109.628 & 135.419 & YBiH$_6$   & Pm-3 \\
abh6-Li-Te  & 1.777 & 734.823 & 90.914  & 107.251 & 128.028 & LiTeH$_6$  & Pm-3 \\
abh6-Hf-Ag  & 6.838 & 187.701 & 41.853  & 106.397 & 146.374 & AgHfH$_6$  & Pm-3 \\
abh6-Ge-Zr  & 2.202 & 586.719 & 85.226  & 106.298 & 122.181 & GeZrH$_6$  & Pm-3 \\
abh6-Ca-Zr  & 2.719 & 495.912 & 81.551  & 106.197 & 114.106 & CaZrH$_6$  & Pm-3 \\
abh6-Sb-Sc  & 1.723 & 749.692 & 90.345  & 105.007 & 126.988 & ScSbH$_6$  & Pm-3 \\
\hline\hline
\end{tabular}
\label{best-screening}
\end{table}

\subsection{Training of EDDPs}

Ephemeral Data Derived Potentials (EDDPs) can be used to accelerate crystal structure prediction. We trained potentials
using the iterative approach described in reference \cite{salzbrenner2023}.

The training starts with \textit{f} single-point-energy calculations of randomly generated structures on which an EDDP is trained. In each iteration, \textit{m} local minima are found by random searching using the current EDDP, and single-point-energies are calculated for 10 ‘shaken‘ structures in the vicinity of each minimum, which are then added to the dataset. PBE and PBEsol functional were used for single-point-energies calculations, with coarse DFT parameters: a 340\,eV plane-wave cutoff, a $k$-point spacing of $2\pi\times~0.05$\,\AA${}^{-1}$, and \textsc{castep} ultrasoft QC5 pseudopotentials.

At the end of each iteration, a new EDDP is trained. For Y--Sb--H potentials, we used \textit{n} cycles iterations and \textit{m} local minima found at pressures randomly chosen between a pressure range. Usually the target pressure should be in the middle of the training pressure range.

The form of the potential is naturally cut-oﬀ at a distance of \textit{rcut}\,(\AA), containing 4 polynomials and a neural network containing a single layer with 20 nodes. The structures in the training data contained between 0--N atoms. By construction, the atoms can come in close contact with a minimal separation distance \textit{minsep} 1-2\,\AA.

We trained several EDDPs, some were trained to target 50\,GPa, and another one was trained to target 120\,GPa. More than 77 000 structures were generated to train 120\,GPa EDDPs, and $\sim$94,000 structures were generated to train 50\,GPa EDDPs. The parameters used to train EDDPs and energies errors are reported in Table~\ref{eddp}.

\begin{table}[htbp]
\centering
\caption{Parameters used for training EDDPs and energies errors in training dataset}
\begin{tabular}{l|ccc}
\hline\hline
Target pressure (GPa) & 50 & 50 & 120  \\
\hline
Pressure range (GPa) & 0-100 & 45-55 & 115--125 \\
\textit{n} cycles & 33 & 31 & 25 \\
\textit{f} random structures & 2000 & 3000 &3000 \\
\textit{m} local minima & 1000 & 3000 &3000 \\
\textit{rcut} (\AA) & 3.75 & 5.4 & 5.4 \\
0--N atoms & Y, Sb=0--5; H= 0--20 & 6--12 & 6--12 \\
Energy window (eV) & 10 & 10 & 10 \\
Functional & PBE & PBEsol & PBEsol  \\
Total generated structures & 28,386 & 94,740 &77,114 \\
number of compositions & 612 &  334 & 334 \\
\hline
RMSE (meV) &  83.36 & 37.39 &37.81 \\
MAE (meV)  & 32.05 & 14.90  &15.68 \\
Spearman &0.99996 & 0.99985 & 0.99996 \\
\hline\hline
\end{tabular}
\label{eddp}
\end{table}

\subsection{Structure searches using EDDPs}

We performed structure searches using EDDPs (to relax the structures) within AIRSS \cite{Pickard2006} and with Hot-AIRSS~\cite{Pickard2025} (to sample lower energies structures).
More than 300,000 structures and 100,000 structures were generated in the searches at 50\,GPa, and 120\,GPa respectively. The single-point energies were calculated after the EDDPs relaxation, using PBEsol functional, with the same coarse parameters as in the training.

\subsection{Molecular Dynamics Simulations}

We have run several molecular dynamics by using the two EDDPs we have trained, to check the kinetic stability of the cubic phase of YSbH$_6$.

A molecular dynamics code, \texttt{ramble} (see ~\cite{salzbrenner2023}), is designed to work well with EDDPs. We used \texttt{ramble} to calculate molecular dynamics trajectories at 50\,GPa and 120\,GPa. Simulation cells were made from cubic supercells which contained 1,000 atoms. We ran NpT calculations at 300\,K, 50\,GPa and 120\,GPa for 0.5\,ns with a timestep of 0.05\,fs, excluding 50\,ps of equilibration.

\begin{figure}[h!]
\centering
\includegraphics[width=0.47\textwidth]{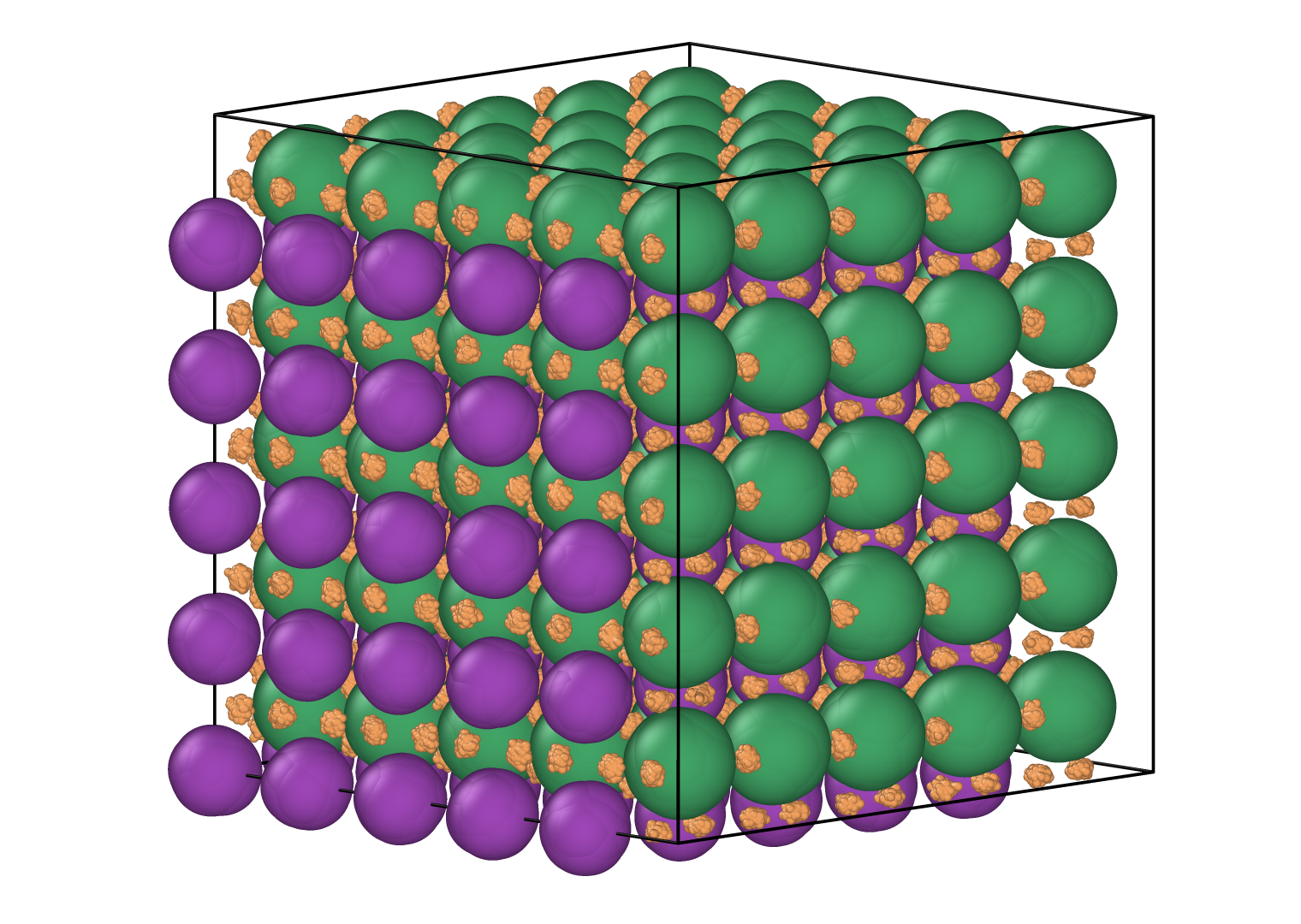}
\caption{Supercell (5$\times$5$\times$5) of YSbH$_{6}$ with 1000 atoms with Sb (purple), Y (green) and H (orange). Hydrogen positions over a 0.5\,ns trajectory (excluding 50\,ps equilibration) at 300\,K and 50\,GPa. }
\label{fig:snapshot}
\end{figure}

\begin{figure}[h!]
\centering
\includegraphics[width=0.7\textwidth]{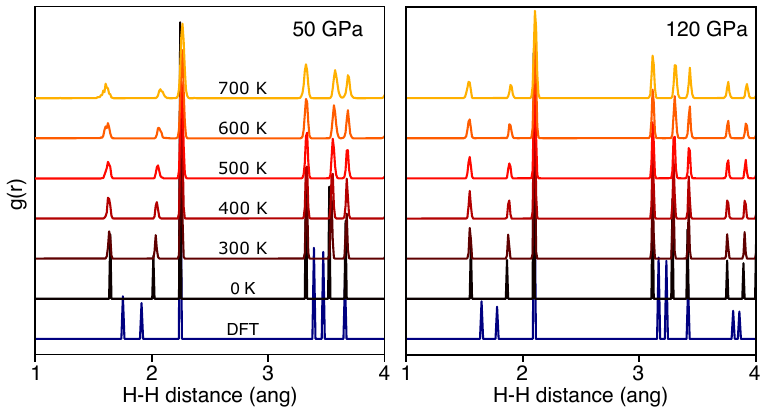}
\caption{Partial radial distribution functions of hydrogen in the YSBH$_6$ obtained from time averaged positions in molecular dynamics and static geometry optimisation (0 K).}
\label{fig:HH}
\end{figure}

Throughout the MD trajectory, at 300\,K the supercell of YSbH$_6$ remains the same at both 50\,GPa and 120\,GPa, with almost no diffusion of H atoms after 0.5\,ns. Figure~\ref{fig:snapshot} shows snapshots at 0.5\,ns of the MD trajectory. The very long time scale of the MD trajectory only achievable by using EDDPs, showcase kinetical stability of the cubic structure at both high and lower pressures (Figure~\ref{fig:HH}).

An interpretation of these results is that the cubic phase of YSbH$_6$ could be synthesised by heating at high pressure 120\,GPa (see section~\ref{synthesis} for possible synthesis route) and followed by quenching to preserve the structure down to lower pressure 50\,GPa.

\clearpage

\section{Calculations of the superconducting critical temperature and dynamic stability}

\subsection{\bf `Coarse' parameters used in DFPT calculations for high-throughput search in A15-type hydrides}

To screen through the set of promising A-15 type structures, we performed coarse DFPT calculations using \textsc{Quantum Epsresso}. The phonon $\mathbf{q}$ grid of $3\times3\times3$ and electron $\mathbf{k}$ mesh of $6\times6\times6$ were used, with the electron phonon matrix elements then interpolated onto a denser $12\times12\times12$ grid. A set of \textsc{GBRV} ultrasoft pseudopotentials were used in this inital screening, since these could be used with a lower plane-wave cutoff energy of 50\,Ry and charge-density cut off of 500\,Ry. The pseudopotentials used the PBE exchange-correlation functional, although subsequent calculations with finer parameters used the PBEsol functional. As seen in Table \ref{best-screening}, the Y-Sb structure was identified as the most promising candidate.

\subsection{`Fine' DFPT parameters and methods for $T_\mathrm{c}$ prediction of YSbH$_6$ }

For the converged $T_{\mathrm{c}}$ predictions, the phonon grid density was increased to $0.06\times2\pi$ \AA$^-1$, corresponding to a $\mathbf{q}$ grid of $5\times5\times5$. The corresponding $\mathbf{k}$ grid was $10\times10\times10$, and the electron phonon matrix elements were interpolated onto a $20\times20\times20$ grid. The \textsc{pseudo-dojo} \textsc{ONCV} pseudopotentials were used, with a kinetic energy cut off of 100\,Ry, to ensure the total energy was converged to within 1 meV/atom. The PBEsol exchange-correlation functional was used for these calculations.

\begin{figure}[ht]
    \centering
    \includegraphics[width=0.7\linewidth]{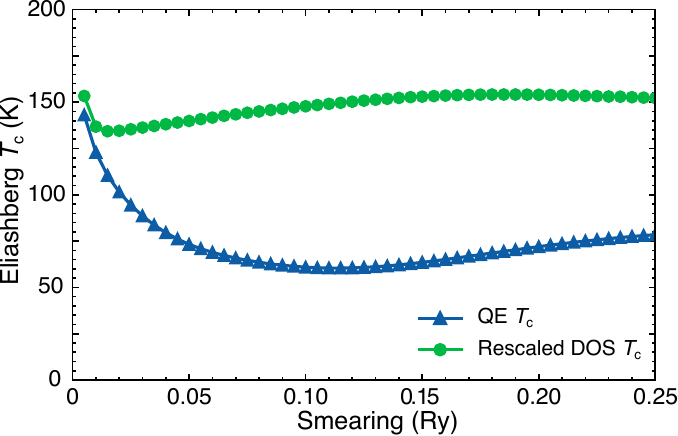}
    \caption{Plot of the calculated Eliashberg $T_{\mathrm{c}}$ against Gaussian smearing width. The constant density of states Eliashberg equations were used, with a Coulomb pseudopotential of $\mu^*=0.227$. The blue curve shows the obtained $T_{\mathrm{c}}$ from \textsc{Quantum Espresso} and \textsc{a2tc}, and the green curve shows the same results with the density of states rescaling applied. The rescaling improves convergence by correcting for errors introduced by a poorly reproduced density of states. }
    \label{fig:Tc_vs_smear}
\end{figure}

The numerical implementation of the electron phonon coupling strength $\lambda$, and subsequently $T_\mathrm{c}$, requires replacing Dirac delta functions with Gaussians of finite width. For underconverged calculations, this can lead to a large spread in the $T_\mathrm{c}$, largely because the density of states at the Fermi energy is not correctly reproduced. To mitigate this, we computed $T_{\rm c}$ over a range of Gaussian broadenings from 0.005--0.250\,Ry, and applied a density of states rescaling. This step involves rescaling the electron phonon spectral function $\alpha^2F(\omega)$, also known as the Eliashberg spectral function, by a factor related to the density of states at the Fermi energy,
\begin{equation}
    \widetilde{\alpha^2F}(\omega, \sigma) =  \frac{\widetilde{N_{\mathrm{F}}}}{N_{\mathrm{F}}(\sigma)} \; \alpha^2F(\omega, \sigma),
    \label{eqn: scaling_factor}
\end{equation}
where $\widetilde{N_{\mathrm{F}}}$ is a high quality density of states calculated with the tetrahedra method, $N_{\mathrm{F}}(\sigma)$ is the density of states calculated with the Gaussian smearing used in the $T_{\mathrm{c}}$ calculation, and $\widetilde{\alpha^2F}$ is the rescaled Eliashberg spectral function. For the coarse calculations, the tetrahedra density of states was calculated on a $18\times18\times18$ grid, and for the fine calculations a $30\times30\times30$ grid was used. A plot of the Eliashberg $T_\mathrm{c}$ against the smearing width is shown in Figure \ref{fig:Tc_vs_smear}.

Two approaches were used to calculate the $T_{\mathrm{c}}$. The first is the semi-empirical Allen-Dynes equation, which is given by \cite{allen1975}
\begin{equation}
    T_{\mathrm{c}}^{\rm{AD}} = f_1 f_2 \times \frac{\omega_{\rm{log}}}{1.2}\exp{\left( -\frac{1.04(1+\lambda)}{\lambda - \mu^*_{\rm{AD}}(1+0.62\lambda)} \right)}.
\end{equation}
Here, $\lambda$ is the electron phonon coupling strength, $\mu^*_{\rm{AD}}$ is the Allen-Dynes Coulomb pseudopotential (set to a value of 0.125), $\omega_{\rm{log}}$ is the logarithmic average of the electron phonon spectral function $\alpha^2F(\omega)$, and $f_1$ and $f_2$ are  pre-factors dependent on $\lambda$, $\mu^*_{\rm{AD}}$, phonon frequencies $\omega$ and the electron phonon spectral function $\alpha^2F(\omega)$. Both $\lambda$ and $\omega_{\rm{log}}$ can be expressed in terms of $\alpha^2F(\omega)$ as
\begin{equation}
    \lambda = \int_0^{\infty} d\omega \frac{2\alpha^2F(\omega)}{\omega} ,
\end{equation}
and
\begin{equation}
    \omega_{\rm{log}} = \exp\left(\frac{2}{\lambda} \int_0^\infty\frac{\alpha^2F(\omega)}{\omega} \ln(\omega) d\omega\right).
\end{equation}

The second approach used to calculate $T_{\mathrm{c}}$ was the Migdal-Eliashberg equations. These were applied with the isotropic approximation, with a Matsubara cut off energy of 7000\,meV. These were solved within the Anderson-Morel Coulomb pseudopotential framework, where the value of $\mu^*_{\rm{ME}}$ was calculated using the expression
\begin{equation}
    \mu^*_{\rm{ME}} = \frac{\mu^*_{\rm{AD}}}  {1 + \mu^*_{\rm{AD}}\ln(\frac{\omega_{\rm{ph}}}{\omega_{\mathrm{C}}})},
\end{equation}
where the Allen-Dynes value $\mu^*_{\rm{AD}}$ was set to 0.125, $\omega_{\rm{ph}}$ is the highest frequency phonon, and $\omega_{\rm{C}}$ is the Matsubara cut off energy. From this, the value was calculated to be 0.227. 

\begin{figure}[ht]
    \centering
    \includegraphics[width=0.7\linewidth]{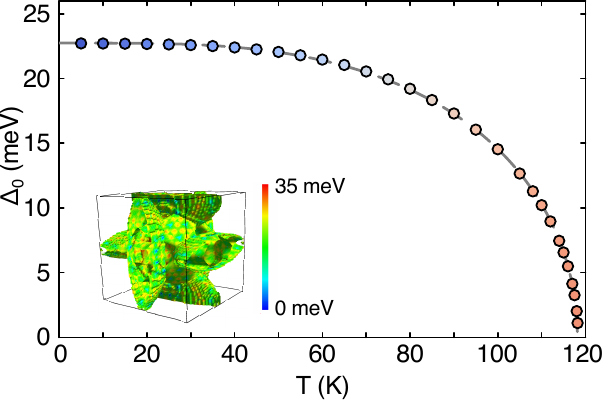}
    \caption{The curve shows the temperature dependence of the superconducting gap $\Delta_0$ calculated with the vDOS isotropic Migdal-Eliashberg equations. The Coulomb pseudopotential was $\mu^*_{\rm{ME}} = 0.227$. The figure in the bottom left shows the value of the superconducting gap for all points on the Fermi surface at a temperature of 50\,K.}
    \label{fig:superconducting_gap}
\end{figure}

The equations were solved in both the constant density of states (cDOS) and variable density of states (vDOS) approximations. In the constant DOS approximate, the value of the density of states is fixed to the value at the Fermi energy and is assumed to be constant within the energy range of interest. In contrast, the variable DOS is able to account for the shape of the DOS, and the Eliashberg equations take the form \cite{Kogler2025}
\begin{equation}
    Z(i \omega_j) =  1 + \frac{T}{N(\epsilon_{\rm{F}})\omega_j} \int d\epsilon\, N(\epsilon) \sum_{j'} \frac{\omega_{j'}Z(i\omega_{j'})}{\Theta(\epsilon, i\omega_{j'})}\lambda(i\omega_j - i\omega_{j'})
\end{equation}
\begin{equation}
    \chi(i\omega_{j}) = - \frac{T}{N(\epsilon_{\rm{F}})}\int d\epsilon \,N(\epsilon) \sum_{j'} \frac{\epsilon - \mu_{\rm{F}} + \chi(i\omega_{j'})}{\Theta(\epsilon, i\omega_{j'})} \lambda(i\omega_{j} - i\omega_{j'})
\end{equation}
\begin{equation}
    \phi(i\omega_j) = \frac{T}{N(\epsilon_{\rm{F}})} \int d\epsilon \,N(\epsilon) \sum_{j'} \frac{\phi(\epsilon, i\omega_{j'})}{\Theta(\epsilon, i\omega_{j'})} \left( \lambda(i\omega_j - i\omega_{j'}) - \mu^*_{\rm{ME}} \right)
\end{equation}

where $\Theta(\epsilon, i\omega_{j'}) = \big(\omega_jZ(i\omega_j)\big)^2 + \big( \epsilon - \mu_{\rm{F}} + \chi(i\omega_j)\big)^2 + \big(\phi(i\omega_j\big)^2) $. The superconducting gap function is then given by $\Delta(i\omega_j) = \frac{\phi(i\omega_j)}{Z(i\omega_j)}$. A plot of the superconducting gap $\Delta_0$ as a function of temperature is shown in Figure \ref{fig:superconducting_gap}. The $T_\mathrm{c}$ is given by the temperature where $\Delta_0$ becomes zero, which occurs at 118\,K.

Lastly, the anisotropic Eliashberg equations were solved using the \textsc{EPW} code \cite{Margine2013,lee2023} to calculate the superconducting gap for all points on the Fermi surface. The $\mathbf{k}/\mathbf{q}$ grid was interpolated from $10^3/5^3$ to $80^3/20^3$. The resulting plot is shown in the inset of Figure \ref{fig:superconducting_gap}. It shows the momentum resolved gap is quite uniform across the Fermi surface, indicating the isotropic approximation used in \textsc{a2tc} and \textsc{IsoME} is a valid approximation.

\clearpage
\subsection{Tuning with pressure}
\begin{figure}[h]
    \centering
    \includegraphics[width=0.5\linewidth]{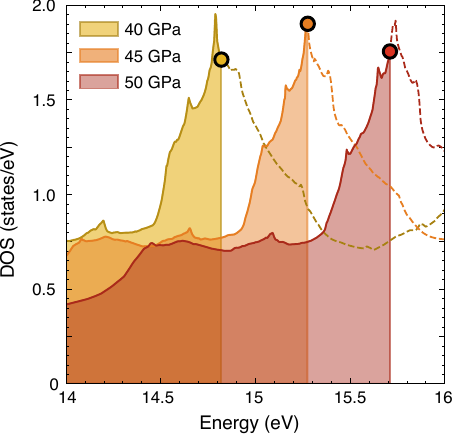}
    \caption{Plot of the density of states for pressures of 40\,GPa, 45\,GPa and 50\,GPa. The density of states at the Fermi energy, and the cDOS estimate of $T_{\rm{c}}$, is maximized at 45\,GPa.}
    \label{fig:dos_vs_press}
\end{figure}

Within the cDOS Eliashberg equations, the electron-phonon coupling strength $\lambda$ is directly proportional to the density of states at the Fermi energy $N_{\rm{F}}$. As shown in Figure~\ref{fig:dos_vs_press}, the application of pressure can alter the position of the density of states, with $N_{\rm{F}}$ reaching a maximum of 1.90\,states/eV at 45\,GPa. There was also a corresponding maximum in the cDOS $T_{\rm{c}}$ of 143\,K. However, the vDOS $T_{\rm{c}}$ remained unchanged, maintaining a value of 118\,K. This illustrates how the sharp peak in the DOS can lead to a discrepancy between the two approaches, with the vDOS less sensitive to small changes in the position of the Fermi level.

\subsection{Dynamic stability}

\begin{figure}[h!]
    \centering
    \includegraphics[width=0.5\linewidth]{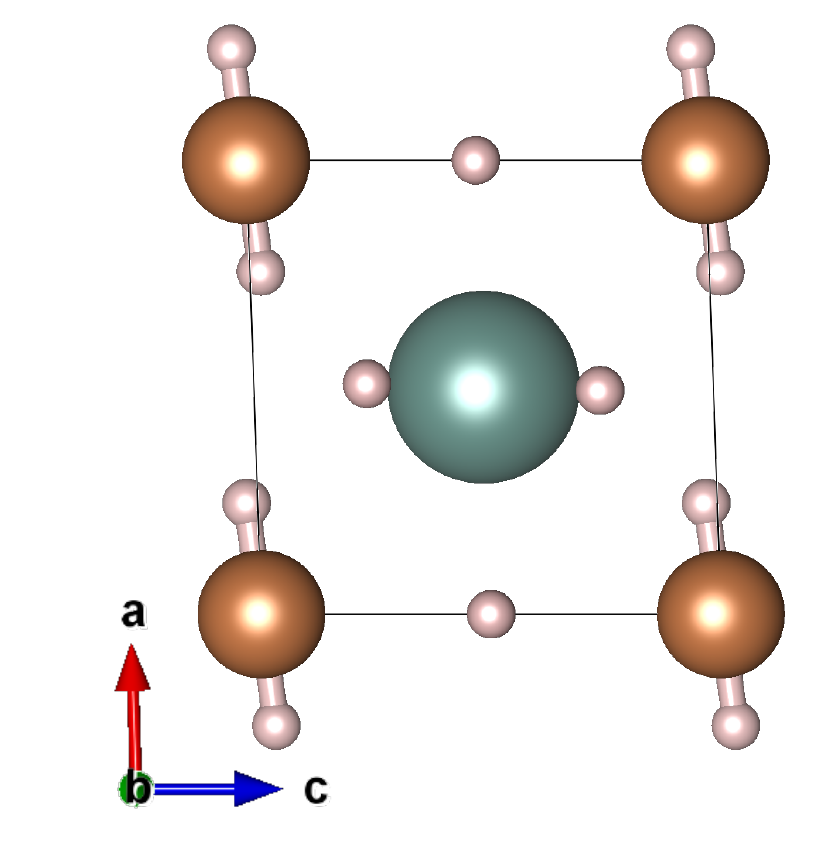}
    \caption{Distorted phase of YSbH$_6$ at 120\,GPa, space group $P2/m$. The system is monoclinic with $\beta = 91.9^{\circ}$ and lattice parameters $a=3.431$\,\AA{}, $b=3.440$\,\AA{}. As the pressure is lowered, this distortion becomes smaller and the system tends towards the cubic structure.}
    \label{fig:distorted}
\end{figure}

Phonon calculations at 120\,GPa  on a $5\times5\times5$ and a $7\times7\times7$ phonon $\mathbf{q}$ grid indicated that the structure was dynamically stable, since no imaginary frequencies were found on these coarse grids. However, Fourier interpolation indicated a shallow imaginary mode at $\mathbf{q} = (0, 0, \frac{1}{10} )$. This is not unexpected, as the interpolation can on occasion result in slightly negative modes around the $\Gamma$ point. To investigate, we performed a direct DFPT calculation at this wavevector. Although using an electronic $\mathbf{k}$ grid of $10\times10\times10$ produced only real frequencies, a more careful calculation on a $20\times20\times20$ electronic grid appeared to confirm the existence of a small imaginary mode. The magnitude of this was found to be -\,13\,cm$^{-1}$ with \textsc{CASTEP}, and -\,23\,cm$^{-1}$ with \textsc{Quantum Espresso}. In principle, this imaginary mode suggests the system is slightly unstable with respect to a small distortion of the hydrogen atoms in a $1\times1\times10$ supercell. Relaxations of this perturbed system led to the discovery of a distorted structure, with very similar enthalpy to the cubic system - within 1\,meV/atom \ref{tab:distorted}. This distorted structure at 120\,GPa is shown in Figure \ref{fig:distorted}. As the pressure was lowered, the distortion and enthalpy difference became smaller, and the system tended towards the cubic structure. The zero-point energy of the cubic and sheared structures at different pressures \ref{tab:ZPE} showcases a difference of less than 2\,meV/atom.

\begin{table}[h]
\centering
\caption{Calculated vDOS $T_{\rm{c}}$ for both the cubic and sheared structures. Both structures here use a $3\times3\times3$ phonon $\mathbf{q}$ point grid, as the low symmetry of the sheared structure made finer grids unfeasible. The sheared structure has a lower $T_{\rm{c}}$, but the value remains high.}

     \begin{tabular}{ccccc}
      \toprule \toprule
       Pressure & \multicolumn{2}{c}{Cubic} &  \multicolumn{2}{c}{Sheared} \\
       
       (GPa)  &    $\lambda$  &  $T_{\rm{c}}$ (K)  &      $\lambda$    &  $T_{\rm{c}}$ (K)     \\
      \midrule
      20 & 3.02 & 122  &  3.01 & 122  \\
      50 & 1.92 & 128  & 1.56  & 110 \\
      120 & 1.31 & 83  & 1.09 & 56 \\
      \bottomrule\bottomrule
      \end{tabular}

\end{table}

\begin{table}[h!]
\centering
\caption{Comparison of the enthalpy, $pV$, and energy of the distorted and cubic structures. These were calculated with \textsc{VASP}~\cite{Kresse1996} and r$^2$SCAN~\cite{Furness2020}. The two systems have very similar enthalpies, with the distorted P2/m being slightly favoured.}
\vspace{1em}
\begin{tabular}{cccccc}
\hline \hline
P (GPa) & spg & H (eV/f.u.) & PV (eV/f.u.) & E (eV/f.u.) \\
\hline
50  & P2/m  & -50.01626 & 15.70080 & -65.71695 \\
100 & P2/m  & -35.58918 & 26.79548 & -62.38456 \\
120 & P2/m  & -30.35524 & 30.68776 & -61.04293 \\
150 & P2/m  & -22.91141 & 36.14540 & -59.05673 \\
200 & P2/m  & -11.35659 & 44.43380 & -55.79032 \\
\hline

50  & Pm-3  & -50.01485 & 15.71318 & -65.72786 \\
100 & Pm-3  & -35.58638 & 26.79754 & -62.38380 \\
120 & Pm-3  & -30.35047 & 30.70336 & -61.05371 \\
150 & Pm-3  & -22.90373 & 36.15723 & -59.06084 \\
200 & Pm-3  & -11.34422 & 44.43802 & -55.78207 \\
\hline \hline
\end{tabular}
\label{tab:distorted}
\end{table}

\begin{table}[h!]
\centering
\caption{Zero-point energies calculated with PBEsol functional with \textsc{Quantum Espresso}~\cite{Giannozzi2020} for cubic and sheared structures. The two energies are very similar as the phonon mode frequencies are largely unchanged.}
\vspace{1em}
\begin{tabular}{lcc}
\hline \hline
P (GPa) & Cubic (eV/atom) & Sheared (eV/atom) \\
\hline
20  & 0.137 & 0.138 \\
50  & 0.171 & 0.170 \\
120 & 0.208 & 0.209 \\
\hline \hline
\end{tabular}
\label{tab:ZPE}
\end{table}

\clearpage
\section{Elastic properties}

Table~\ref{tab:Elastic-20} and table~\ref{tab:Elastic-50} show that cubic YSbH$_6$ is elastically stable at 20 and 50\,GPa because  $C_{44} > 0$. However, table~\ref{tab:Elastic-120} shows that cubic YSbH$_6$ is mechanically unstable at 120\,GPa because  $C_{44} < 0$ .

\begin{table}[h!]
    \centering
    \caption{Elastic constants C$_{ij}$ for cubic YSbH$_6$ at 20\,GPa, are calculated using ref.~\cite{Walker2012} and \textsc{castep}  with a strain of 0.01. Bulk modulus (B$^v$: Voigt, B$^r$: Reuss, B$^{vrh}$: Hill) and shear modulus (G$^v$: Voigt, G$^r$: Reuss, G$^{vrh}$: Hill) are also calculated. }
    \vspace{1em}
    \begin{tabular}{cc} \hline \hline
Elastic constant C$_{ij}$ & GPa\\
\hline 
 c$_{11}$ & 203.6$\pm$1.5 \\ 
c$_{12}$ & 124.4$\pm$0.6 \\ 
c$_{13}$ & 124.4$\pm$0.6 \\ 
c$_{22}$ & 203.6$\pm$1.5 \\ 
c$_{23}$ & 124.4$\pm$0.6 \\ 
c$_{33}$ & 203.6$\pm$1.5 \\ 
c$_{44}$ &  44.7$\pm$8.3 \\ 
c$_{55}$ &  44.7$\pm$8.3 \\ 
c$_{66}$ &  44.7$\pm$8.3 \\ 
 & \\ 
B$^v$ & 150.8$\pm$0.3 \\ 
B$^r$ & 150.8$\pm$1.9 \\ 
B$^{vrh}$ & 150.8$\pm$1.1 \\ 
G$^v$ &  42.7$\pm$2.9 \\ 
G$^r$ &  42.5$\pm$2.6 \\ 
G$^{vrh}$ &  42.6$\pm$2.7 \\ 
\hline \hline
    \end{tabular}
    \label{tab:Elastic-20}
\end{table}

\begin{table}[h!]
    \centering
    \caption{Elastic constants C$_{ij}$ for cubic YSbH$_6$ at 50\,GPa, are calculated using ref.~\cite{Walker2012} and \textsc{castep} with a strain of 0.01. Bulk modulus (B$^v$: Voigt, B$^r$: Reuss, B$^{vrh}$: Hill) and shear modulus (G$^v$: Voigt, G$^r$: Reuss, G$^{vrh}$: Hill) are also calculated.}
    \vspace{1em}
    \begin{tabular}{cc} \hline \hline
Elastic constant C$_{ij}$ & GPa\\
\hline 
c$_{11}$ & 270.0$\pm$5.4 \\ 
c$_{12}$ & 232.3$\pm$1.9 \\ 
c$_{13}$ & 232.3$\pm$1.9 \\ 
c$_{22}$ & 270.0$\pm$5.4 \\ 
c$_{23}$ & 232.3$\pm$1.9 \\ 
c$_{33}$ & 270.0$\pm$5.4 \\ 
c$_{44}$ &   2.7$\pm$10.6 \\ 
c$_{55}$ &   2.7$\pm$10.6 \\ 
c$_{66}$ &   2.7$\pm$10.6 \\ 
 & \\ 
B$^v$ & 244.8$\pm$1.2 \\ 
B$^r$ & 244.8$\pm$64.6 \\ 
B$^{vrh}$ & 244.8$\pm$32.9 \\ 
G$^v$ &   9.1$\pm$3.7 \\ 
G$^r$ &   4.1$\pm$8.5 \\ 
G$^{vrh}$ &   6.6$\pm$6.1 \\ 
\hline \hline
    \end{tabular}
    \label{tab:Elastic-50}
\end{table}

\begin{table}[h!]
    \centering
    \caption{Elastic constants C$_{ij}$ for cubic YSbH$_6$ at 120\,GPa, are calculated using ref.~\cite{Walker2012} and \textsc{CASTEP} with a strain of 0.01. Bulk modulus (B$^v$: Voigt, B$^r$: Reuss, B$^{vrh}$: Hill) and shear modulus (G$^v$: Voigt, G$^r$: Reuss, G$^{vrh}$: Hill) are also calculated.}
    \vspace{1em}
    \begin{tabular}{cc} \hline \hline
Elastic constant C$_{ij}$ & GPa\\
\hline 
c$_{11}$ & 501.1$\pm$3.9 \\ 
c$_{12}$ & 446.8$\pm$1.3 \\ 
c$_{13}$ & 446.8$\pm$1.3 \\ 
c$_{22}$ & 501.1$\pm$3.9 \\ 
c$_{23}$ & 446.8$\pm$1.3 \\ 
c$_{33}$ & 501.1$\pm$3.9 \\ 
c$_{44}$ & -39.7$\pm$12.9 \\ 
c$_{55}$ & -39.7$\pm$12.9 \\ 
c$_{66}$ & -39.7$\pm$12.9 \\ 
 & \\ 
B$^v$ & 464.9$\pm$0.8 \\ 
B$^r$ & 464.9$\pm$81.0 \\ 
B$^{vrh}$ & 464.9$\pm$40.9 \\ 
G$^v$ & -12.9$\pm$4.5 \\ 
G$^r$ & -2523.8$\pm$18081.4 \\ 
G$^{vrh}$ & -1268.4$\pm$9043.0 \\ 
\hline \hline
    \end{tabular}
    \label{tab:Elastic-120}
\end{table}

\clearpage

\section{EDDPs search and convex hull at 120\,GPa}

\begin{figure}[h!]
\centering
\includegraphics[width=0.9\textwidth]{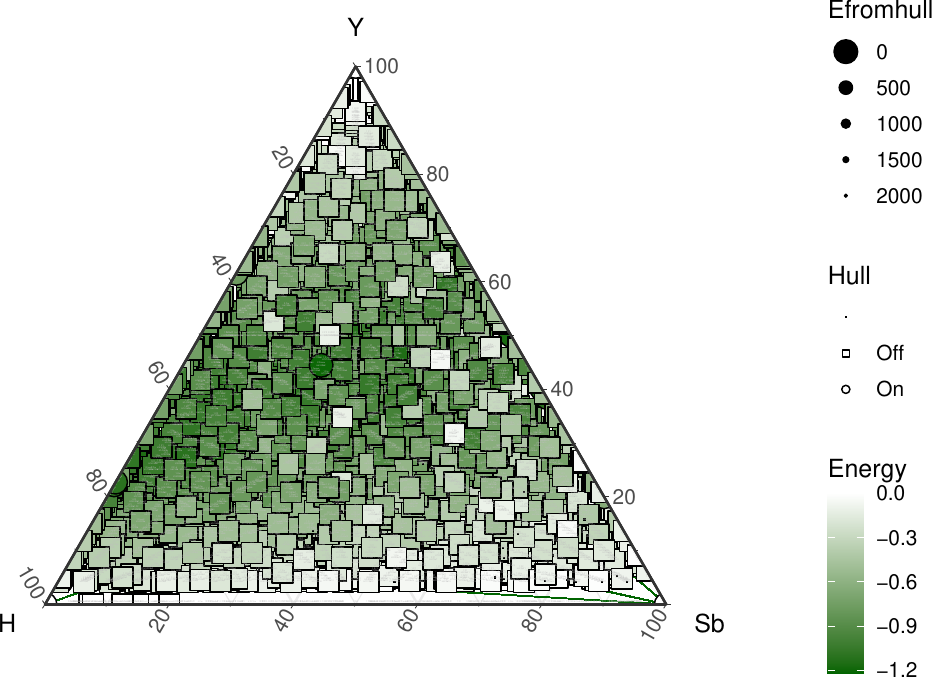}
\caption{Y--Sb--H ternary phase diagram of one search done with EDDPs at 120\,GPa. Structures are relaxed with EDDPs and only the single point energy is calculated with \textsc{CASTEP}. The convex hull with DFT relaxed structures is presented in the article in Figure 2. }
\label{fig:search}
\end{figure}

\begin{table*}[h!]
\centering
\caption{Y--Sb--H phases lying on the convex hull at 120 GPa. Enthalpy per formula unit $H$ (eV/f.u.), formation enthalpies $\Delta H$$_{\rm form}$ (eV/at) and energy distances from the hull ($E_{\rm h}$ (eV/at)) of Y--Sb--H phases on the convex hull at 120\,GPa as given by cryan.}
\begin{tabular}{lcccccccc}
\hline\hline
Formula & $P$ (GPa) & $V$ (\AA$^3$/f.u.) & $H$ (eV/f.u.) & $\Delta H_{\rm form}$ (eV/at) & $E_{\rm h}$ (eV/at) & nfu & Space group \\
\hline
H$_2$Y$_2$Sb    & 120.01 &  47.788 &  -4535.165 & -1.292 &  0.000 & 1 &  I4/mmm   \\
H$_4$Y$_2$Sb    & 120.03 &  51.534 &  -4564.193 & -1.275 &  0.000 & 2 &  I4/mcm   \\
YSb             & 120.00 &  29.428 &  -3445.901 & -1.223 &  0.000 & 1 &  P4/mmm   \\
YSb$_2$         & 120.05 &  44.589 &  -5831.889 & -1.222 &  0.000 & 2 &  Fddd     \\
HY$_5$Sb$_3$    & 120.04 & 118.892 & -12471.853 & -1.204 &  0.000 & 1 &  Cmmm     \\
H$_3$Y          & 119.98 &  20.590 &  -1103.237 & -1.177 &  0.000 & 1 &  Fm-3m    \\
HY$_3$Sb$_6$    & 120.06 & 135.225 & -17509.602 & -1.165 &  0.000 & 1 &  Imm2     \\
Y$_3$Sb$_2$     & 120.02 &  72.613 &  -7951.368 & -1.154 &  0.000 & 2 &  C2/m     \\
HY$_2$Sb$_4$    & 119.97 &  90.746 & -11677.651 & -1.132 &  0.000 & 1 &  C2       \\
HY$_3$Sb        & 120.07 &  59.515 &  -5579.643 & -1.107 &  0.000 & 1 &  Pm-3m    \\
H$_2$Y$_9$Sb$_3$& 120.00 & 175.901 & -16724.001 & -1.068 &  0.000 & 1 &  Cmmm     \\
HY$_6$Sb$_2$    & 120.00 & 116.414 & -11144.344 & -1.045 &  0.000 & 1 &  P4/mmm   \\
HYSb$_2$        & 119.98 &  46.241 &  -5845.664 & -1.040 &  0.000 & 2 &  Aem2     \\
HY              & 120.02 &  15.908 &  -1073.945 & -0.989 &  0.000 & 1 &  Fm-3m    \\
H$_4$Y          & 120.01 &  21.321 &  -1116.705 & -0.979 &  0.000 & 1 &  I4/mmm   \\
Y$_3$Sb         & 119.94 &  57.305 &  -5564.635 & -0.952 &  0.000 & 1 &  Pm-3m    \\
H$_7$Y          & 120.00 &  27.346 &  -1156.808 & -0.644 &  0.000 & 2 &  C2/c     \\
Y$_7$Sb         & 119.99 & 111.622 &  -9799.421 & -0.481 &  0.000 & 1 &  I4mm     \\
H$_{14}$Y       & 119.75 &  42.154 &  -1249.846 & -0.348 &  0.000 & 1 &  P-1      \\
HY$_{18}$       & 119.99 & 245.127 & -19071.944 & -0.122 &  0.000 & 1 &  P-1      \\
H$_6$Sb         & 119.99 &  26.985 &  -2464.630 & -0.025 &  0.000 & 2 &  P2$_1$/m \\
H               & 119.99 &   2.034 &    -13.281 &  0.000 &  0.000 & 8 &  C2/c     \\
Sb              & 119.97 &  15.397 &  -2384.769 &  0.000 &  0.000 & 1 &  Im-3m    \\
Y               & 119.99 &  13.544 &  -1058.686 &  0.000 &  0.000 & 1 &  Fm-3m    \\
\hline\hline
\label{totalenergy}
\end{tabular}
\end{table*}

\clearpage

\section{Convex hull at 50\,GPa}

\begin{figure}[h!]
\centering
\includegraphics[width=0.9\textwidth]{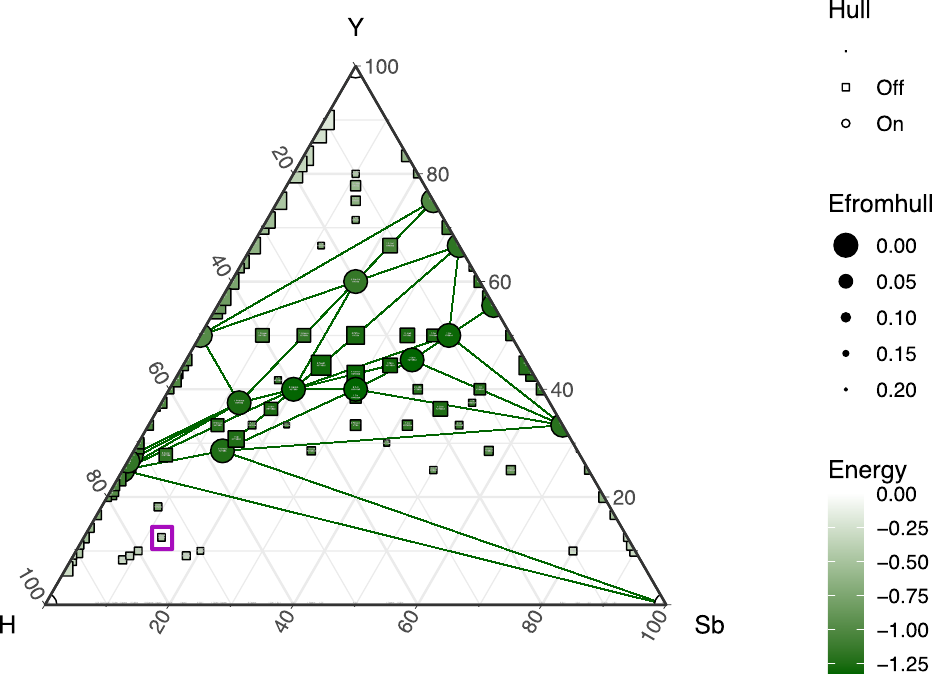}
\caption{Ternary convex hull at 50\,GPa, calculated with PBEsol functional. Dark green circles indicate the thermodynamically stable phases, solid lines are ridges connecting thermodynamically stable points. Metastable phases are shown as squares, their size are smaller according to their energy distance from the convex hull (eV/at). The convex hull s 180 structures collected from different searches, with a total of 146 different compositions. Structures showed on the plot have an energy distance from the hull E$_h$\,$\leq$\,206\,meV/at. The color scale is according the formation enthalpy (eV/at) of each structure relative to Y, Sb and H elements. The purple square indicates metastable $Pm\Bar{3}$ YSbH$_6$.}
\label{fig:ternary}
\end{figure}

\begin{table}[h!]
\centering
\caption{Y--Sb--H phases on the convex hull at 50 GPa. Enthalpy per formula unit $H$ (eV/f.u.), formation enthalpies $\Delta H$$_{\rm form}$ (eV/at) and energy distances from the hull ($E_{\rm h}$ (eV/at)) of Y--Sb--H phases on the convex hull at 120\,GPa as given by cryan}
\begin{tabular}{lcccccccc}
\hline \hline
Formula & $P$ (GPa) & $V$ (\AA$^3$/f.u.) & $H$ (eV/f.u.) & $\Delta H_{\rm form}$ (eV/at) & $E_{\rm h}$ (eV/at) & nfu & Space group \\
\hline
H$_3$Y$_4$Sb$_3$ & 50.00 & 135.062 & -11493.999 & -1.321 &  0.000 & 1 & R-3    &  \\
H$_2$Y$_5$Sb$_4$ & 50.13 & 167.698 & -14938.418 & -1.318 &  0.000 & 1 & C2/m   &  \\
H$_2$Y$_2$Sb     & 50.02 &  58.664 &  -4558.083 & -1.313 &  0.000 & 2 & I4/mmm &  \\
HY$_5$Sb$_4$     & 50.00 & 166.159 & -14922.571 & -1.298 &  0.000 & 1 & Cm     &  \\
H$_4$Y$_3$Sb     & 50.01 &  82.508 &  -5655.617 & -1.256 &  0.000 & 1 & P4/mmm &  \\
Y$_5$Sb$_4$      & 50.01 & 162.936 & -14906.508 & -1.250 &  0.000 & 1 & C2/m   &  \\
H$_4$Y$_2$Sb     & 49.99 &  62.661 &  -4588.831 & -1.235 &  0.000 & 3 & I4/mcm &  \\
YSb$_2$          & 49.98 &  54.027 &  -5853.153 & -1.201 &  0.000 & 2 & C2/m   &  \\
Y$_2$Sb          & 50.03 &  54.173 &  -4526.379 & -1.175 &  0.000 & 2 & Fddd   &  \\
H$_3$Y           & 49.99 &  24.938 &  -1113.066 & -1.171 &  0.000 & 2 & Fm-3m  &  \\
H$_{11}$Y$_4$    & 50.01 &  98.959 &  -4436.579 & -1.158 &  0.000 & 1 & C2/c   &  \\
H$_{13}$Y$_5$    & 50.04 & 122.678 &  -5533.949 & -1.150 &  0.000 & 1 & I4/m   &  \\
HY$_3$Sb         & 50.05 &  73.554 &  -5608.296 & -1.145 &  0.000 & 2 & Pm-3m  &  \\
Y$_3$Sb          & 50.04 &  71.184 &  -5592.265 & -1.006 &  0.000 & 2 & Pm-3m  &  \\
HY               & 49.98 &  19.979 &  -1081.660 & -0.970 &  0.000 & 2 & Fm-3m  &  \\
Sb               & 49.99 &  18.548 &  -2392.081 &  0.000 &  0.000 & 2 & Im-3m  &  \\
H                & 50.01 &   2.920 &    -14.332 &  0.000 &  0.000 & 3 & I4/mmm &  \\
Y                & 49.99 &  18.121 &  -1065.386 &  0.000 &  0.000 & 2 & P-6m2  &  \\
\hline \hline
\end{tabular}
\end{table}

\clearpage

\section{Kinetic stability}

We use the partial radial distribution function to characterise the hydrogen network in the time averaged structure at a range of temperatures and pressures. All MD calculations were conducted with the \texttt{ramble} code from the EDDP software suite. In each case a 0.5 ns trajectory in the NPT ensemble was used with a 0.5 fs time-step. The 0 K line shown in black is generated from a geometry optimization with an EDDP using the \texttt{repose} code, the DFT line is generated from a geometry optimization in \textsc{CASTEP}. Small discrepancies are visible between the DFT and EDDP relaxed structures however these correspond to distances that are fractions of an Angstrom and are unlikely to have a significant impact on T$_c$. Broadening in the peaks visible at elevated temperatures shows increased hydrogen mobility.
\clearpage

\section{Bonding and charge}

\subsection{Electron localisation function (ELF)}

\begin{figure}[h!]
\centering
\includegraphics[width=0.8\textwidth]{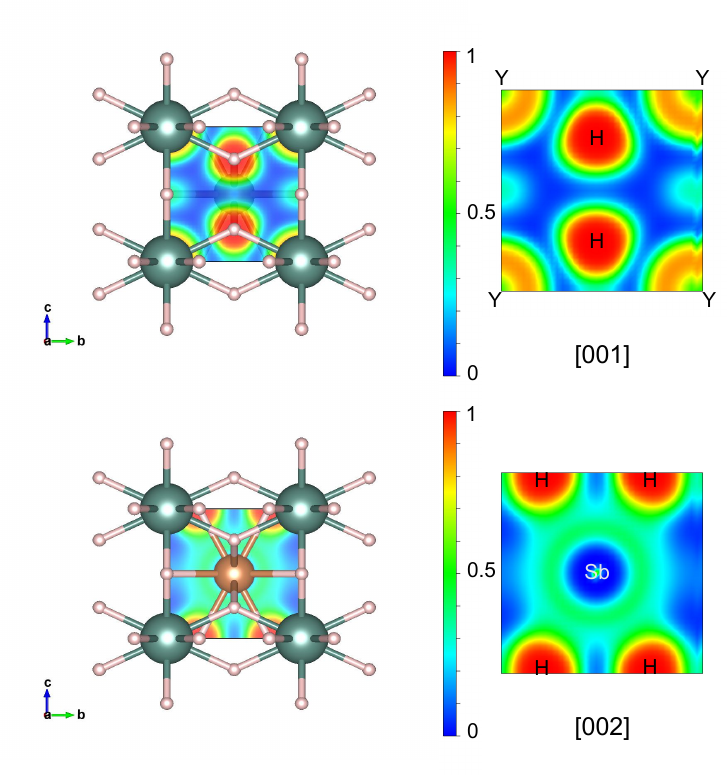}
\caption{The calculated ELF of $Pm\Bar{3}$ YSbH$_6$ at 50\.GPa.}
\label{elf}
\end{figure}
\clearpage

\subsection{Topological analysis of the electron density}

In \ref{table:topo1} and \ref{table:topo2} we present the results from the topological analysis of the electron density of YSbH$_6$, calculated at 50 GPa, based on the Quantum Theory of Atoms in Molecules (QTAIM) \cite{Bader1990}. In \ref{table:topo1}, report the Bader charges and volumes for each atomic species. In \ref{table:topo2}, instead, we report several topological descriptors for the analysis of chemical bond. Specifically, we show the electron density $\rho$(\textbf{r}) and the Laplacian $\nabla^2$$\rho$(\textbf{r}) at the critical points, together with the kinetic, potential, and total energy densities at the critical points, $G$(\textbf{r}), $V$(\textbf{r}) and $H$(\textbf{r}), as well as the ration between the potential and the kinetic energy densities $|V|$(\textbf{r})$/$$G$(\textbf{r}) \cite{Gatti2005}. 

Atomic charges and volumes \ref{table:topo1} quantify the partial charge transfer from Y and Sb to the hydrogen atoms, which become partially negatively charged. At the same time, the heavy elements assume almost identical atomic volumes.
On the other hand, the bonding analysis \ref{table:topo2}, quantify the stronger bonds between H -- Y and H -- Sb compared to the H -- H interactions. Topologically, H -- Sb appear more covalent in character, and stronger in energy, compared to the other two. Yet, all of them present the topological characteristics of metal-metal or donor-acceptor interactions, which are intermediate to the closed-shells and shared-shells \cite{Gervasio2004}.

\begin{table}[h!]
\centering
\caption{Bader charges and volumes.}
\begin{tabular}{lcc}
\hline \hline
Atom & Charge (e) & Volume (\AA$^3$) \\
\hline
H   & -0.35  & 3.65 \\
Y   &  0.99  & 13.69 \\
Sb  &  1.11  & 13.65 \\

\hline \hline
\end{tabular}
\label{table:topo1}
\end{table}

\begin{table}[h!]
\centering
\caption{Chemical bonding analysis}
\begin{tabular}{lccccccc}
\hline \hline
Bond & $\rho$(\textbf{r}) (e$/$\AA$^3$) & $\nabla^2$$\rho$(\textbf{r}) (e$/$\AA$^5$) & $G$(\textbf{r}) (eV$/$\AA$^3$) & $V$(\textbf{r}) (eV$/$\AA$^3$) & $H$(\textbf{r}) (eV$/$\AA$^3$) & $|V|$(\textbf{r})$/$$G$(\textbf{r}) \\
\hline
H -- Sb  &  0.41  &  1.16  & 6.44  & -10.68  & -4.24  & 1.66 \\
H -- Y   &  0.43  &  3.40  & 9.76  & -13.05  & -3.29  & 1.34 \\
H -- H   &  0.29  &  1.74  & 5.01  & -6.70   & -1.69  & 1.34 \\
\hline \hline
\end{tabular}
\label{table:topo2}
\end{table}

\clearpage
\section{Possible synthesis routes}
\label{synthesis}

Our calculations show that the decomposition of YSbH$_6$ is thermodynamically favoured at 120 GPa ($\Delta H$ = -627.810\,meV/f.u. , $\Delta H$ = -26.16\,meV/atom), following the reaction

\begin{equation}
\mathrm{3YSbH_6} \xrightarrow[]{} \mathrm{YH_4} + \mathrm{2YH_7} + \mathrm{3Sb}
\label{eq:reaction1}
\end{equation}

However, the low enthalpy of decomposition also suggests that YSbH$_6$ might be synthesized as metastable phase or as a kinetic product following different paths of reaction. Therefore, here we propose two possible synthetic routes to obtain YSbH$_6$ in diamond anvil cell (DAC).

One possibility would be to use YH$_3$ (CAS:13598-57-7) and SbH$_3$ (CAS:7803-52-3) as reactants. On one hand, YH$_3$ is a solid, semiconducting, black powder that would be easy to handle and load in a DAC. On the other, SbH$_3$ is a toxic gas with a melting point of $\sim$ 185 K, that must be handled with care. However, gas loading techniques are now widely used in HP syntheses and the combination of YH$_3$ and SbH$_3$ would give a very good control on the stoichiometry of the reaction. If the loading of was successful, the following reaction might promoted under high-pressure, and perhaps, high-temperature (laser heating) conditions:

\begin{equation}
\mathrm{YH_3} + \mathrm{SbH_3} \xrightarrow[\text{HP}]{\Delta} \mathrm{YSbH_6}
\label{eq:reaction1}
\end{equation}

The formation enthalpy for this reaction is favourable with a $\Delta H_\text{form}$ = -137\,meV/f.u. at 120 GP.

\vspace{1em}
An alternative route to obtain YSbH$_6$ in DAC would be to use yttrium antimonide, YSb, (CAS: 12186-97-9) and hydrogen gas loading. YSb is a solid, small band gap, semiconductor, also obtainable as a fine powder. Hydrogen gas, instead, would need more care due to its flammability. On one hand, hydrogen gas might be easier to handle and load in a DAC compared to SbH$_3$, on the other, the combination of YSb with H$_2$ would not guarantee the same control on the stoichiometry of the reaction as in the previous case.

\begin{equation}
\mathrm{YSb} + 3\mathrm{H_2} \xrightarrow[\text{HP}]{\Delta} \mathrm{YSbH_6}
\label{eq:reaction2}
\end{equation}

However, as for the previous case, the reaction might be promoted under laser heating conditions. However, in this case, the formation enthalpy is very large, $\Delta H_\text{form}$ = -2411\,meV/f.u. at 120 GPa.

\clearpage

\section{Crystal structure of YSbH$_6$ at 50\,GPa (.res format)}
If one wants to use the .res file, one should directly download the .res file and not use the displayed .res format below.
\begin{lstlisting}
TITL YSbH6-2021941-527-117 49.9909 49.268794 -3.54727773E+003 0 0  8 (Pm-3) n - 1
REM CASTEP 25.1 from code version 325614284+ default Fri Apr 19 14:02:46 2024 +0100
REM Functional PBE for solids (2008) Relativity Koelling-Harmon Dispersion off
REM Cut-off 900.0000 eV Grid scale 2.0000 Gmax 30.7390 1/A FBSC none
REM Offset 0.000 0.000 0.000 No. kpts 119 Spacing 0.02
REM Total runtime: 536.2 s
REM Overall efficiency: 96 %
REM
REM ./YSbH6.cell (d8b6dcbb55eb849cf5adb739a50f26e2)
REM ## AIRSS Version 0.9.4 June 2023 commit v0.9.4-109-g9759ecccc979b07271
REM
REM
REM cmdline: crud.pl -mpinp 4
REM
REM
REM Y 3|2.0|7|8|9|40U:50:41:42
REM Sb 3|2.5|2.5|1.5|7|8|9|50:51:42U2.2U2.2
REM H 1|0.6|13|15|17|10(qc=8)
REM
CELL 1.54180    3.66598    3.66598    3.66598   90.00000   90.00000   90.00000
LATT -1
SFAC H  Y  Sb
H      1  1.0000000000000  0.5000000000000  0.7604649132217 1.0  mu:-0.300
H      1  0.2395350867783  1.0000000000000  0.5000000000000 1.0  mu:-0.300
H      1  0.7604649132217  1.0000000000000  0.5000000000000 1.0  mu:-0.300
H      1  1.0000000000000  0.5000000000000  0.2395350867783 1.0  mu:-0.300
H      1  0.5000000000000  0.2395350867783  1.0000000000000 1.0  mu:-0.300
H      1  0.5000000000000  0.7604649132217  1.0000000000000 1.0  mu:-0.300
Y      2  1.0000000000000  1.0000000000000  1.0000000000000 1.0  mu:0.773
Sb     3  0.5000000000000  0.5000000000000  0.5000000000000 1.0  mu:1.029
END
\end{lstlisting}

\bibliographystyle{unsrt}
\bibliography{biblio.bib}